\documentclass[%
 preprint,
 amsmath,amssymb,
 aps, physrev,
]{revtex4-2}

\usepackage{physics}
\usepackage{graphicx}
\graphicspath{ {./figures/} }
\usepackage[colorlinks=true, linkcolor=blue, citecolor=blue, urlcolor=blue]{hyperref}
\usepackage[caption=false]{subfig}
\usepackage{multirow}
\usepackage{booktabs}
\usepackage{gensymb}

\begin{document}

\preprint{APS/123-QED}

\title{Two-Mode Bosonic State Tomography with Single-Shot Joint Parity Measurement of a Trapped Ion}

\author{Honggi Jeon \footnote{Present Address: Duke Quantum Center, Duke University, Durham, North Carolina 27701, USA}}
\thanks{These two authors contributed equally}
\affiliation{Department of Computer Science and Engineering, Seoul National University}
\affiliation{Automation and Systems Research Institute, Seoul National University}

\author{Jiyong Kang}
\thanks{These two authors contributed equally}
\affiliation{Department of Computer Science and Engineering, Seoul National University}
\affiliation{Automation and Systems Research Institute, Seoul National University}

\author{Wonhyeong Choi}
\affiliation{Department of Computer Science and Engineering, Seoul National University}
\affiliation{Automation and Systems Research Institute, Seoul National University}

\author{Kyunghye Kim}
\affiliation{Department of Computer Science and Engineering, Seoul National University}
\affiliation{Automation and Systems Research Institute, Seoul National University}

\author{Jaehun You}
\affiliation{Department of Computer Science and Engineering, Seoul National University}
\affiliation{Automation and Systems Research Institute, Seoul National University}

\author{Taehyun Kim}
\email{taehyun@snu.ac.kr}
\affiliation{Department of Computer Science and Engineering, Seoul National University}
\affiliation{Automation and Systems Research Institute, Seoul National University}
\affiliation{Inter-university Semiconductor Research Center, Seoul National University}
\affiliation{Institute of Computer Technology, Seoul National University}
\affiliation{Institute of Applied Physics, Seoul National University}
\affiliation{NextQuantum, Seoul National University}

\date{\today}

\begin{abstract}
The full characterization of a continuous-variable quantum system is a challenging problem. For the trapped-ion system, a number of methods of measuring the quantum states have been developed, including the measurement of the Q quasi-probability function and the density matrix elements in the Fock basis, but these approaches are often slow and difficult to scale to multi-mode states. Here, we demonstrate a novel and powerful scheme for measuring a continuous-variable quantum state that uses the direct single-shot measurement of the joint parity of the phonon states of a trapped ion. We drive a spin-dependent bichromatic beam-splitter interaction that coherently exchanges phonons between different harmonic oscillator modes of the ion. This interaction encodes the joint parity information into the relative phase between the two spin states, enabling measurement of the combined phonon-number parity across multiple modes in a single shot. Leveraging this capability, we directly measure multi-mode Wigner quasi-probability distributions to perform quantum state tomography of an entangled coherent state, and calculate various quantum informational quantities with a model-based estimation of the density matrix. We further show that the single-shot joint parity measurement can be used to detect parity-flip errors in real time. By post-selecting the parity measurement outcomes, we experimentally demonstrate the partial recovery of coherence, effectively implementing an error mitigation technique. Lastly, we identify the various sources of error affecting the fidelity of the spin-dependent beam-splitter operation and study the feasibility of high fidelity operations. The interaction studied in this work can be extended to more than two modes, and is highly relevant to continuous-variable quantum computing and quantum metrology.
\end{abstract}

\maketitle

\section{Introduction}

Continuous-variable quantum states (CVQS) are a promising platform for quantum computation, simulation, communication and sensing. They have been used to encode qubit states using such encoding schemes as the cat code or the Gottesman--Kitaev--Preskill (GKP) code~\cite{Munro2000, goettesman2001, matsos_2024, de_neeve_error_2022, wang_schrodinger_2016}. The cat code has advantages such as hardware-efficient error correction and the ability to exploit strongly biased noise—where phase-flip errors occur much more frequently than bit-flip errors~\cite{Guillaud2019, Putterman2025Hardware}. The GKP code encodes quantum information in the periodic superposition in phase space. It allows one to detect and correct errors using low-overhead operations such as spin-dependent displacements in trapped ion implementations~\cite{fluhmann_encoding_2019, Matsos2025Universal}. These continuous variable encodings provide an alternative approach towards fault-tolerant quantum computing. There have also been proposals for hybrid quantum computation schemes, which use both the discrete (qubits) and continuous (bosonic modes) degrees of freedom~\cite{Liu2024, Lee2024, Omkar2020}. They have also been utilized to simulate quantum chemistry using analog quantum simulation~\cite{shen_quantum_2018, Whitlow2023Quantum, So2024, MacDonell2021Analog, Sun2025Quantum}, and as quantum sensors~\cite{Burd2019, Gilmore2021Quantum, Metzner2024, Birrittella2021}. 

Despite their usefulness, CVQS are not easy to characterize. Their inherently large Hilbert space requires a substantial number of measurements, and efficient characterization often relies on access to a suitable interaction Hamiltonian, which may not always be experimentally available. This challenge becomes even more severe for entangled CVQS, where one has to find a suitable multi-mode interaction to enable efficient characterization. In some platforms such as the circuit quantum electrodynamics (QED) system, a dispersive interaction Hamiltonian involving the superconducting qubit and the microwave resonator exists. It has been used to encode the parity of the bosonic mode in the qubit state~\cite{Sun2014Tracking, Blais2021}. The full characterization or quantum state tomography (QST) of multimode bosonic states, therefore has been realized only in a handful of experiments on platforms such as circuit QED systems, flying photons, and trapped ions~\cite{wang_schrodinger_2016, Sun2014Tracking,  ourjoumtsev_preparation_2009, Ra2019Non, wang_flying_2022, Besse2020, Hacker2019Deterministic, Zhubing2022, Valahu2023Direct, Whitlow2023Quantum}.

The trapped-ion system, owing to its excellent controllability and coherence times, has been widely used as a platform for realizing CVQS~\cite{kienzler_observation_2016, Monroe1996, Meekhof1996}. The quantum state tomography of CVQS has been realized in a number of trapped-ion experiments using the Q function, characteristic function, and direct measurement of density matrix elements~\cite{Leibfried1996, fluhmann_encoding_2019, Fluhmann2020Direct, Zhubing2022, Millican2025}. Notably, characteristic function QST was used in Refs.~\cite{fluhmann_encoding_2019, Fluhmann2020Direct} and was successfully expanded to two-mode CVQS in Ref.~\cite{Valahu2023Direct, Whitlow2023Quantum, Millican2025}. Also, the density matrices of multi-mode quantum states were reconstructed in Ref.~\cite{Zhubing2022} by observing the phonon-number--dependent Rabi oscillation of multiple product spin basis states, which yields the elements of the density matrix in Fock state basis. However, this scheme requires the individual detection and control of multiple spins in an ion chain, which increases the experimental overhead.

In this work, we present the experimental realization of a single-shot parity measurement of multi-mode bosonic states encoded in the motional degrees of freedom of a trapped ${}^{171}\text{Yb}^+$ ion. This is enabled by a spin-dependent beam splitter (SBS) operation which we realize by driving a bichromatic interaction with detunings of $\pm(\omega_1 - \omega_2)$ with respect to the carrier transition, where $\omega_{1}$, $\omega_{2}$ are the angular frequencies of the two radial motional modes. Our realization of multi-mode CVQS characterization uses the single-ion Jaynes--Cummings Hamiltonian which is native to most trapped-ion systems. This Hamiltonian can be implemented in other platforms that support a Jaynes--Cummings-type spin--boson interactions across multiple modes, such as trapped neutral atoms and superconducting qubit--cavity systems~\cite{Blais2021, Kaufman2012, Lu2022}. The SBS interaction enables a non-destructive and single-shot measurement of the joint parity, an important characteristic of bosonic states. It is non-destructive, as the Fock state information is preserved under a deterministic, trackable mode swap and phase-space rotation.

It is known that the Wigner function is the expectation value of the displaced phase-space inversion operator, or the boson number parity operator, up to normalization~\cite{Royer_1977}. The Wigner function of a quantum state contains the full information about the state and can be mapped to the density matrix using matrix inversion. An efficient reconstruction technique for multi-mode bosonic states that is based on Wigner function was recently developed as well~\cite{Kevin2024Efficient}. Moreover, parity is an important observable in quantum metrology, where it can saturate the Cramér--Rao bound~\cite{Birrittella2021, Joo2011}, and in continuous-variable quantum computing (CVQC), where it serves as a stabilizer operator~\cite{Ofek2016Extending,Gertler2021Protecting,Chamberland2022}. Despite these advantages, a multi-mode Wigner function tomography has not been realized thus far in trapped-ion systems.

We note that this interaction was experimentally realized in a previous work~\cite{nguyen2021}, where the spin-dependent beam splitter (SBS) interaction was used to measure the overlap integral between two motional quantum states rather than to measure the joint parity. In another experiment, an SBS interaction was realized in $\sigma_z$ basis with the $\ket{\uparrow}$ and $\ket{\downarrow}$ states as the control states. This interaction was used to implement a phononic network for boson sampling~\cite{Chen2023Scalable}. A similar interaction was used to measure the parity of a single mode in another work~\cite{Gan2020}. This interaction uses a running optical lattice with an oscillating polarization, which modulates the Stark shift that the $\ket{S_{1/2}, F=1, m_F=1}$ state of $^{171}\text{Yb}^+$ ion experiences. This method is incompatible with the magnetically insensitive hyperfine qubit state that is widely used.

This paper is organized as follows. In section~\ref{sect:theory}, we introduce the spin-dependent beam splitter interaction Hamiltonian and derive the time evolution, showing that it encodes the multi-mode parity information in the spin state. We also describe how the Wigner function can be obtained in a trapped-ion system by combining displacement operations with parity measurements. In section~\ref{sect:exp_results}, we describe the experimental setup, characterize the interaction, and carry out Wigner function QST on two exemplary quantum states; two-mode Fock states and entangled coherent states (ECSs). Then we carry out a model-based estimation of the density matrix of the ECS, and calculate quantum informational quantities from it. Lastly, we demonstrate that the single-shot and non-destructive nature of the parity measurement can be used to detect and mitigate the effects of heating error with post-selection even after it has decohered significantly. 

\section{Theory}\label{sect:theory}

\subsection{Spin-dependent Beam Splitter Interaction}
The spin-dependent beam splitter (SBS) interaction used in our experiment is realized by a bichromatic Hamiltonian with two frequencies that are detuned by the frequency difference of the two radial motional modes of a trapped ion, $\pm(\omega_1-\omega_2)$, with respect to the carrier transition, where $\omega_1$ and $\omega_2$ are the mode frequencies of the radial modes $1$ and $2$, respectively. Throughout the rest of the paper, we will assume $\omega_2>\omega_1$. In the interaction picture that factors out the qubit and harmonic oscillator parts of the Hamiltonian, $H_0=\frac{\hbar\omega_\mathrm{HF}}{2}\sigma_z + \hbar\omega_1 (a_1^\dagger a_1 + \frac{1}{2}) + \hbar\omega_2 (a_2^\dagger a_2 + \frac{1}{2})$, where $\omega_\mathrm{HF}$ denotes the hyperfine qubit energy, the SBS interaction is described as follows:
\begin{equation}\label{eq:ham-sbs}
    \begin{split}
        H_\text{SBS}=-\frac{\hbar g}{2}(\sigma_+ a_1^{\dagger}a_2e^{i\phi_r}+\sigma_- a_1a_2^{\dagger}e^{-i\phi_r} + \sigma_+ a_1a_2^{\dagger}e^{i\phi_b}+\sigma_- a_1^{\dagger}a_2e^{-i\phi_b}) \\
        = -\frac{\hbar g}{2}(a_1^{\dagger}a_2e^{-i\phi_M}+a_1a_2^{\dagger}e^{i\phi_M})\sigma_{\phi_S},
    \end{split}
\end{equation}
where $g=\eta_1\eta_2\Omega$ denotes the interaction strength, with $\eta_1$ and $\eta_2$ being the Lamb--Dicke parameters for each motional mode and $\Omega$ the Rabi frequency of the carrier transition. As this interaction is second-order, the interaction strength is weaker than the carrier transition by factor of $\eta_1 \eta_2$. The operators $a_1, a_1^{\dagger}$ and $a_2, a_2^{\dagger}$ are the bosonic annihilation and creation operators for modes $1$ and $2$, respectively. $\sigma_+$ and $\sigma_-$ are the spin raising and lowering operators and $\sigma_{\phi_S}$ is the spin operator in $\phi_S$ basis, defined as $\sigma_{\phi_S}=(\sigma_+ e^{i\phi_S} + \sigma_- e^{-i\phi_S})$. The spin and motional phases are defined as $\phi_{S}=(\phi_b + \phi_r)/2$ and $\phi_{M}=(\phi_b - \phi_r)/2$, where $\phi_r$ and $\phi_b$ are the phases of the red- and blue-detuned laser tones that drive the bichromatic Hamiltonian. For the remainder of the paper, we will define $\sigma_0=\sigma_x$ where $\phi_S=0$ without loss of generality. Now we consider the time evolution operator of $H_\text{SBS}$:
\begin{equation}\label{eq:unit-sbs}
    U_\text{SBS}(t)=\exp\left(-\frac{iH_\text{SBS}t}{\hbar}\right)=\exp\left(i\frac{\theta}{2}\sigma_x \left(a_1^{\dagger}a_2e^{-i\phi_M}+a_1a_2^{\dagger}e^{i\phi_M}\right)\right),
\end{equation}
where $\theta=g t$. From here on, we will denote $U_\text{SBS}(t)$ as $U_\text{SBS}$ for notational simplicity unless otherwise specified. We can immediately see that when the spin state is $\ket{\pm}=(\ket{\downarrow}\pm\ket{\uparrow})/\sqrt2$, the eigenstates of $\sigma_x$, Eq.~(\ref{eq:unit-sbs}) is identical to the unitary operator for beam splitting interaction with a tunable phase $\phi_M$ and angle $\theta$ whose sign depends on the spin~\cite{Campos1989, Leonhardt_2003, nguyen2021}.
To understand how the motional state of a trapped ion transforms under the spin-dependent beam splitter interaction, we examine the transformation of the creation operator using the Schwinger representation of the angular momentum operators $J_x=(a_1a_2^{\dagger}+a_1^{\dagger}a_2)/2$ and $J_y=(a_1a_2^{\dagger}-a_1^{\dagger}a_2)/(2i)$~\cite{schwinger_angular_1965}. With $J_x$ and $J_y$, Eq.~(\ref{eq:unit-sbs}) can be expressed as $U_\text{SBS}=\exp(i \theta \sigma_x J_{\phi_M})$ where $J_{\phi_M}=J_x\cos{\phi_M}+J_y\sin{\phi_M}$. It follows that
\begin{equation}\label{eq:creation_op_transform}
    \begin{split}
        U_\text{SBS} (a_1^\dagger) U_\text{SBS}^\dagger = a_1^\dagger \cos(\theta\sigma_x/2) + i a_2^\dagger e^{-i\phi_M} \sin(\theta\sigma_x/2), \\
        U_\text{SBS} (a_2^\dagger) U_\text{SBS}^\dagger = i a_1^\dagger e^{i\phi_M} \sin(\theta\sigma_x/2) + a_2^\dagger \cos(\theta\sigma_x/2).
    \end{split}
\end{equation}
Therefore, for a quantum state of a trapped ion with its spin $\ket{\pm}$ and two-mode Fock state $\ket{n_1}\ket{n_2}$, we can show that
\begin{equation*}\label{eq:time_sbs}
    \begin{split}
        U_\text{SBS}(t)\ket{\pm}\ket{n_1}\ket{n_2} = \frac{(U_\text{SBS}(a_1^{\dagger})U_\text{SBS}^{\dagger})^{n_1}(U_\text{SBS}(a_2^{\dagger})U_\text{SBS}^{\dagger})^{n_2}}{\sqrt{n_1!n_2!}}\ket{\pm}\ket{0}\ket{0} \\
        =\frac{(\cos(g t/2)a_1^{\dagger}\pm ie^{-i\phi_M}\sin(g t/2)a_2^{\dagger})^{n_1}}{\sqrt{{n_1}!}}
         \frac{(\cos(g t/2)a_2^{\dagger}\pm ie^{i\phi_M}\sin(g t/2)a_1^{\dagger})^{n_2}}{\sqrt{{n_2}!}}\ket{\pm}\ket{0}\ket{0}.
    \end{split}
\end{equation*}

\subsection{Single-Shot Joint Parity Measurement \label{sect:single-shot-joint-parity-measurement}}

For an initial wave function $\ket{\downarrow}\ket{n_1}\ket{n_2}=1/\sqrt{2}(\ket{+}+\ket{-})\ket{n_1}\ket{n_2}$, the spin-dependent beam splitter $\pi$-pulse---corresponding to an interaction time $t_\pi=\pi/(\eta_1\eta_2\Omega)$---transforms it into the following wave function:
\begin{equation}\label{eq:parity-sbs}
    U_\text{SBS}\left(\frac{\pi}{\eta_1\eta_2\Omega}\right)\ket{\downarrow}\ket{n_1}_1\ket{n_2}_2=\frac{i^{n_1+n_2}e^{-i(n_1-n_2)\phi_M}}{\sqrt{2}}(\ket{+}+(-1)^{n_1+n_2}\ket{-})\ket{n_2}_1\ket{n_1}_2.
\end{equation}
Therefore, when the spin state is measured in the $\sigma_z$ basis, the outcome will be $\ket{\uparrow}$ ($\ket{\downarrow}$) if the joint parity of the two-mode Fock state is odd (even). In the case of a superposition of two-mode Fock states, the spin-dependent beam splitter interaction generates spin--motion entanglement: the $\ket{\uparrow}$ state becomes entangled with the odd-parity component, and the $\ket{\downarrow}$ state with the even-parity component. Upon spin-state detection, the wave function collapses to either an even or an odd phonon-number parity state, conditional on the measurement outcome. This corresponds to a single-shot joint parity measurement, and the population of the measured spin state is directly mapped to the expectation value of the joint parity operator.

We note that, however, the operation in Eq.~(\ref{eq:parity-sbs}) does not constitute the canonical form of the non-destructive joint parity operator $\mathcal{P}\equiv\exp(i\pi (a_1^\dagger a_1 + a_2^\dagger a_2))$, as it swaps the Fock numbers and introduces additional phase factors between the Fock states. The global phase in Eq.~(\ref{eq:parity-sbs}) becomes a relative phase between different Fock components, and the interaction results in a motional phase rotation of the form
\begin{equation}\label{eq:motion-phase-rotation}
    \exp \left( i  \left( \frac{\pi}{2}-\phi_M \right) n_1 \right) \exp \left( i \left( \frac{\pi}{2} + \phi_M \right) n_2 \right).
\end{equation}
By tuning the motion phase $\phi_M$, the rotation angle applied to each mode can be adjusted. In particular, when $\phi_M=0$, both modes experience a $\pi/2$-rotation in the phase space. This rotation is deterministic and can be calibrated out.

\subsection{Wigner Function Measurement}
The Wigner function of a motional state can be measured using parity measurements combined with displacements in phase space \cite{Royer_1977}. For a two-mode motional state, the Wigner function is
$W(\beta_1, \beta_2) = \frac{4}{\pi^2} \expval{D_1(\beta_1)D_2(\beta_2)\mathcal{P}D_1(-\beta_1)D_2(-\beta_2)}$,
where $D_1(\beta_1)$ and $D_2(\beta_2)$ are the displacement operators acting on the modes $1$ and $2$, displacing them by $\beta_1$ and $\beta_2$, respectively. This implies that the Wigner function $W(\beta_1, \beta_2)$ is directly proportional to the expectation value of the displaced joint parity operator $\mathcal{P}$.

The displacement operators can be implemented by applying a bichromatic spin-dependent force (SDF)~\cite{kienzler_observation_2016, Jeon2024Experimental}, which is a first-order interaction in terms of Lamb--Dicke parameters, whose Hamiltonian is
\begin{equation} \label{eq:hamiltonian-sdf}
    H_\text{SDF}^{j}=\frac{\hbar\eta_{j}\Omega}{2}\sigma_{\phi_S + \frac{\pi}{2}}(a_je^{-i\phi_M}+a_j^{\dagger}e^{i\phi_M}),
\end{equation}
where $j=1, 2$ denotes the motional mode affected by the force, and $a_j$ and $a_j^\dagger$ are the annihilation and creation operators associated with that mode. The time-evolution operator corresponding to Eq. (\ref{eq:hamiltonian-sdf}), when applied for a duration of $t$, becomes the spin-dependent displacement operator $D_j(\sigma_x\alpha_j) \equiv \exp\left(\sigma_x (\alpha_j a_j^\dagger - \alpha_j^* a_j)\right)$ where the phase-space displacement vector is defined as $\alpha_j=-i\eta_j\Omega e^{i\phi_M}t/2$, assuming $\phi_S=-\pi/2$ for simplicity.

\begin{figure}
    \centering
    \includegraphics[width=0.8\linewidth]{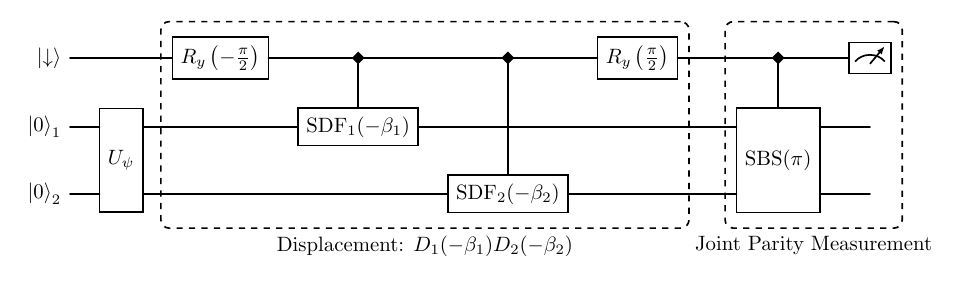}
    \caption{\label{fig:wigner-sequence-diagram} Experimental sequence for the two-mode Wigner function measurement. After initializing the spin to $\ket{\downarrow}$ and preparing the target state $\ket{\psi}_{12}=U_\psi\ket{0}_1\ket{0}_2$, coherent displacements---$\operatorname{SDF}_1(-\beta_1)$ and $\operatorname{SDF}_2(-\beta_2)$---are applied in the respective phase spaces to probe the Wigner function $W(\beta_1, \beta_2)$. The diamond-shaped control symbols represent the spin-dependence in $\sigma_x$. The spin-dependent beam splitter $\pi$-pulse, $\operatorname{SBS}(\pi)$, then maps the joint parity information onto the spin state. Finally, the expectation value of the joint parity operator is extracted from the spin-state population, and the result is rescaled by a factor of $4/\pi^2$ to obtain the Wigner function value. The diagram was generated using Quantikz~\cite{Alastair2018}.}
\end{figure}

However, since this interaction is spin-dependent, the spin state must be prepared in an eigenstate of $\sigma_x$ to realize a pure displacement operator $D_j(\alpha_j)$. This is achieved by applying $\pi/2$ rotations of the spin about the $y$ axis of the Bloch sphere before and after the SDF interaction, as shown in Fig.~\ref{fig:wigner-sequence-diagram}. Using these operations, one can measure the Wigner function $W(\beta_1, \beta_2)$ of an arbitrary two-mode motional state, assuming the spin is prepared in $\ket{\downarrow}$.

\subsection{Extension beyond Two Motional Modes}\label{sect:multimode-extension}
\begin{figure}
    \includegraphics[width=0.7\textwidth]{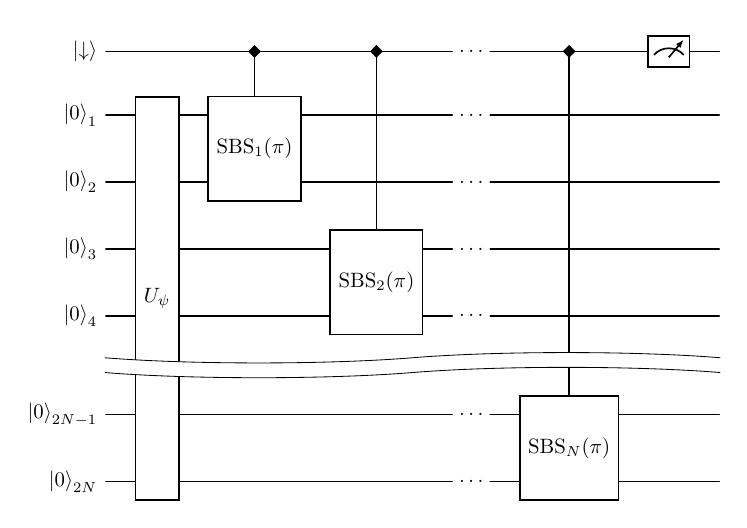}
    \caption{\label{fig:multimode-extension} Circuit diagram representing the multimode joint-parity measurement scheme. $U_\psi$ prepares a $2N$-mode bosonic state, and each spin-dependent beam splitter $\pi$-pulse, $\operatorname{SBS}_j(\pi)$, encodes the joint parity of the mode pair $(2j-1, 2j)$ onto the spin.}
\end{figure}
The same procedure can be extended to more than two motional modes. The relative spin phase accumulated in the $\ket{-}$ state of Eq.~(\ref{eq:parity-sbs}) corresponds to the joint parity of the mode pair $(1, 2)$. When there are multiple disjoint mode pairs, by sequentially applying spin-dependent beam splitter $\pi$-pulses to each mode pair, the joint parities of all pairs are coherently accumulated into the spin phase, as illustrated in Fig.~\ref{fig:multimode-extension}. As a result, the final spin state becomes $\ket{\uparrow}$ ($\ket{\downarrow}$) when the total joint parity---defined as the product of the pairwise joint parities---is odd (even). This enables measurement of the joint parity across multiple modes. If the number of modes is odd, an auxiliary mode can be initialized in the vacuum state to pair with the remaining unpaired mode. The $2N$-mode Wigner function can be generalized as
\begin{equation}
    W(\boldsymbol{\beta})=\left(\tfrac{2}{\pi}\right)^{2N} \expval{\prod_{j=1}^{2N}{D_j(\beta_j)} \prod_{k=1}^{2N}{\mathcal{P}_k} \prod_{l=1}^{2N}{D_l(-\beta_l)}},
\end{equation}
where $\boldsymbol{\beta}\equiv(\beta_1, \beta_2, \dots, \beta_{2N})$ and $\mathcal{P}_k\equiv\exp \left(i\pi a_k^\dagger a_k\right)$. This can be measured by the multimode joint parity readout after displacement operators $D_l(-\beta_l)$.

\section{Experimental Results}\label{sect:exp_results}
In this section, we provide various experimental results of the SBS interaction. First, we apply the beam splitter interaction to two-mode Fock states and observe the time evolution of the spin state in Sec.~\ref{sect:characterization-sbs}.
Then, we use the interaction to generate the quantum state tomograms of two types of two-mode bosonic states: two-mode Fock states in Sec.~\ref{sect:characterization-sbs} and ECS in Sec.~\ref{sect:tomography-ecs}. The results are compared with the simulation, and from the tomography results we estimate the density matrix by fitting a model that considers various error sources to the data. The density matrix is then used to calculate various quantum informational quantities such as the fidelity, purity, and eigenvalues of the experimentally generated state. Finally, we demonstrate the recovery of quantum coherence from a decohered quantum state using single-shot joint parity measurement and post-selection in Sec.~\ref{sect:recovery-parity-selection}.

\subsection{Experimental Setup}

\begin{figure}
    \centering
    \includegraphics[width=0.5\linewidth]{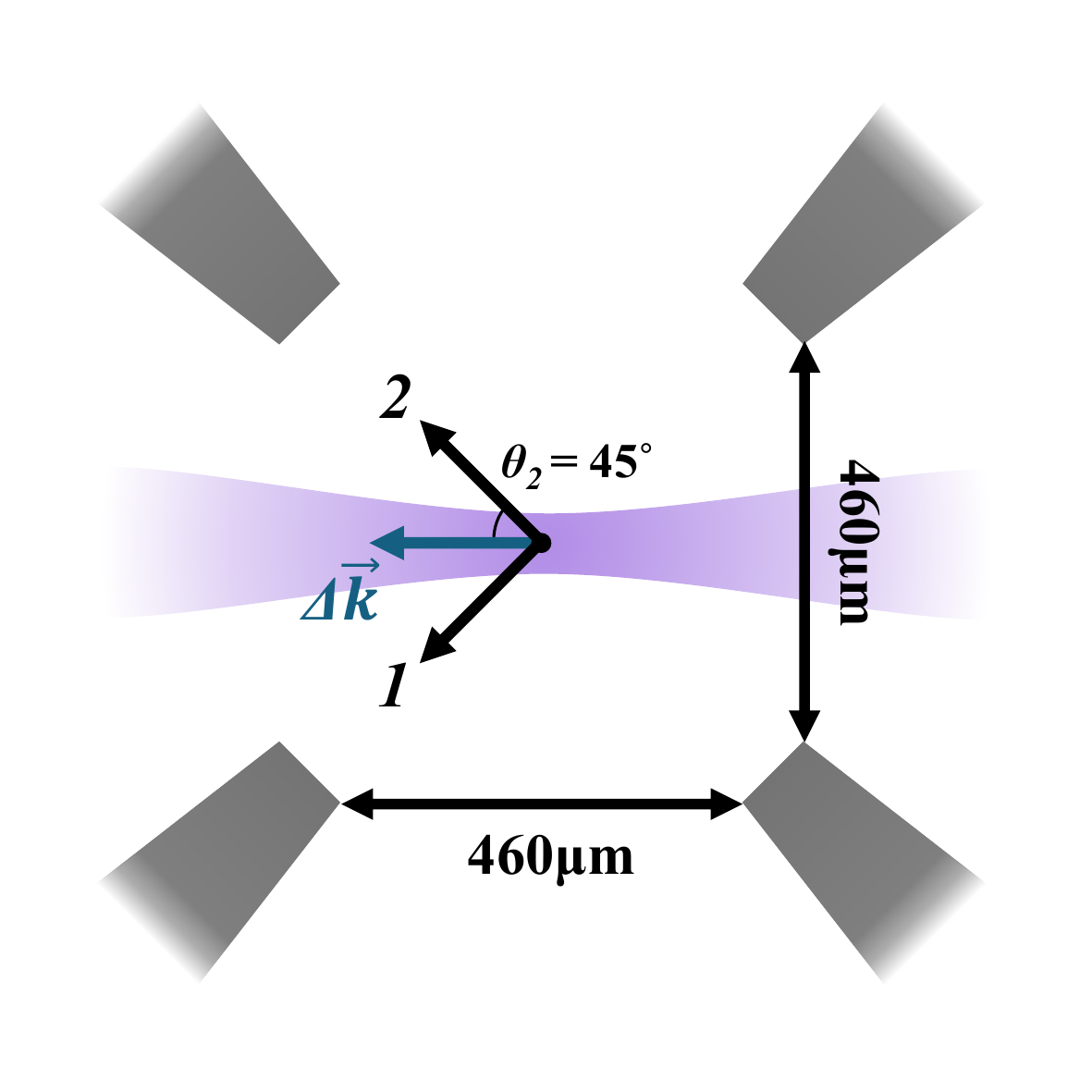}
    \caption{\label{fig:radial-mode-axes} Schematic diagram of the radial RF and ground electrodes along with the principal axes of the radial motional modes $1$ and $2$. Each opposing electrode pair is shorted. A DC offset is applied to the RF electrode pair aligned with the mode $1$, while the other pair is grounded. The vector $\Delta k$ indicates the momentum kick imparted by the Raman transition.}
\end{figure}

We trap a single $^{171}\text{Yb}^+$ ion using a macroscopic linear Paul trap composed of blade-shaped radial RF electrodes and a pair of axial end-cap DC electrodes as shown in Fig.~\ref{fig:radial-mode-axes}~\cite{Jeon2024Experimental}. The qubit state is encoded in the hyperfine levels of the $S_{1/2}$ manifold: $\ket{\downarrow}\equiv \ket{F=0,\,m_F=0}$ and $\ket{\uparrow}\equiv \ket{F=1,\,m_F=0}$. Under typical experimental conditions, the secular frequencies of the radial pseudopotential and axial DC potential are approximately 1~MHz and 118~kHz, respectively. To lift the degeneracy of the radial modes, we apply a DC offset to the RF voltage, resulting in mode frequencies of $\omega_1 / (2\pi) \approx 910\ \mathrm{kHz}$ and $\omega_2 / (2\pi) \approx 1270\ \mathrm{kHz}$. A large radial mode splitting, $(\omega_2 - \omega_1) / (2\pi) \approx 360\ \mathrm{kHz}$, helps reduce the undesired coupling of the interaction laser to the axial mode and the carrier transition when driving the spin-dependent beam splitter interaction, which uses bichromatic beams symmetrically detuned by the splitting. To improve motional coherence and long-term stability of secular frequencies, the RF voltage amplitude is actively stabilized using a servo controller incorporating a capacitive voltage divider, a rectifier, and an RF mixer~\cite{Jeon2024Experimental, Johnson2016Active}. The measured motional coherence time is 2--4 ms; the dephasing mechanism appears inhomogeneous (see Appendix~\ref{sect:error/finite-motional-coherence-time}). The radial modes are sideband-cooled to $\bar n \approx 0.03$. The measured motional heating rates are 5--6 and 17--39 quanta/s for mode~1 and 2, respectively, with day-to-day drift.

We use a 355-nm pulsed laser with acousto-optic modulators (AOMs) to drive Raman transition. To compensate for fluctuations in the laser repetition rate, we implement a feed-forward stabilization system for the beat-note lock~\cite{Mount2016Scalable}. The beam is split into two paths, which are focused onto the trapped ion and arranged to propagate perpendicularly to each other. This configuration yields Lamb--Dicke factors of $\eta_1=0.10$ and $\eta_2=0.085$ for the two radial modes.

\subsection{Characterization of the Spin-dependent Beam Splitter Interaction \label{sect:characterization-sbs}}

\subsubsection{Time Evolution of Spin State \label{sect:time-evolution}}

To verify the correct operation of the SBS interaction, we observe the time evolution of the spin state during the interaction, after preparing two-mode Fock states in the motional modes. The experiment starts by initializing the two motional modes and the spin state via sideband cooling and optical pumping. A two-mode Fock state is then prepared by applying blue- and red-sideband $\pi$-pulses in an alternating sequence. If the target phonon number is odd, a carrier $\pi$-pulse is applied after the final blue-sideband pulse to return the spin to $\ket{\downarrow}$. After the phonon state preparation, the spin state is measured to confirm that it remains in $\ket{\downarrow}$. Experiments yielding $\ket{\uparrow}$ at this step are discarded to increase the fidelity of the process.

When the spin-dependent beam splitter interaction is applied to a two-mode Fock state where one mode contains $n$ phonons and the other is in the vacuum state, the population of the $\ket{\uparrow}$ state evolves in time as 
\begin{equation}
    P_{\uparrow}(t)=\sum_{k=odd}^{n} \binom{n}{k}\cos^{2(n-k)}\left(\frac{g t}{2}\right)\sin^{2k}\left(\frac{g t}{2}\right).
\end{equation}

\begin{figure}
    \centering
    \includegraphics[width=0.4\linewidth]{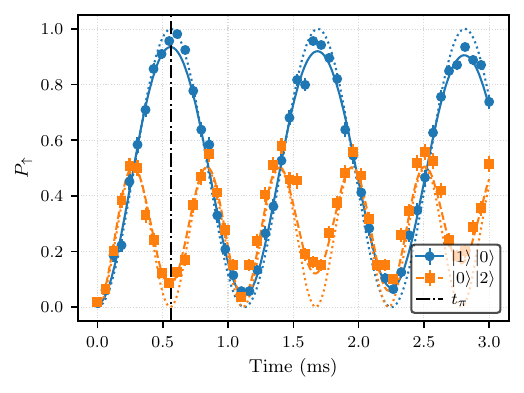}
    \caption{\label{fig:beam-splitter-time-evolution} Oscillation of spin population under spin-dependent beam splitter interaction for Fock states $\ket{1}\ket{0}$ and $\ket{0}\ket{2}$. Scatter points show experimental data, while the solid and dashed curves represent simulations including motional heating. Error bars reflect quantum projection noise from 300 repetitions per data point. Dotted curves illustrate the noiseless simulation. The black dash-dotted line indicates $t_\pi$, at which the joint parity is mapped onto the spin population.}
\end{figure}

The experimentally observed time evolution of two two-mode Fock states is shown in Fig.~\ref{fig:beam-splitter-time-evolution}. The data are overlaid with the numerical simulation results which include imperfect initialization and heating of the motional modes. The simulation also includes the red- and blue-sideband pulses that prepare the two-mode Fock states. It assumes an initial mean phonon number of 0.03 for both modes and heating rates of 5.3 and 17.0 quanta/s for the modes $1$ and $2$, respectively (see Appendix~\ref{sect:error/heating-rate} for details). The theoretical curves were generated by a Lindblad master equation simulation that takes into account the Rabi frequency, initial phonon numbers, and heating rates for both modes. We used the parameters as they were measured in separate experiments except for the Rabi frequency which tended to change presumably because of beam pointing drift. To determine the Rabi frequency, we fitted the simulation to the experimental data, and obtained $2\pi \times 103.0 \ \mathrm{kHz}$ and $2\pi \times 104.5 \ \mathrm{kHz}$ for the states $\ket{1}\ket{0}$ and $\ket{0}\ket{2}$, respectively. The measured $\ket{\uparrow}$ population is rescaled to account for state-detection infidelity, as described in Appendix~\ref{sect:error/qubit-detection}. The simulation and the experimental data show a good agreement, with root mean square errors (RMSEs) of 0.031 and 0.043 for the states $\ket{1}\ket{0}$ and $\ket{0}\ket{2}$, respectively. Especially, at $0.55 \ \mathrm{ms}$, corresponding to the $\pi$ time of the interaction, the spin state evolves to $\ket{\uparrow}$ for the odd joint parity state ($\ket{1}\ket{0}$) and to $\ket{\downarrow}$ for the even joint parity state ($\ket{0}\ket{2}$).

\subsubsection{Wigner Function Measurement \label{sect:fock-wigner-measurement}}

Next, we measure the Wigner functions of the two-mode Fock states $\ket{1}\ket{1}$ and $\ket{2}\ket{1}$. Since Fock states are eigenstates of the number operators $a_j^\dagger a_j$, they acquire only a global phase under single-mode phase-space rotations $\exp \left(i\theta a_j^\dagger a_j\right)$. Their single-mode Wigner functions are therefore rotationally symmetric. Consequently, for two-mode Fock states, any two-dimensional Wigner tomogram constructed with one axis drawn from mode~1 and the other from mode~2 is identical. To demonstrate two-mode quantum state tomography, for each Fock state we take a single two-dimensional Wigner tomogram whose axes are the real axes in the phase space of the two radial modes.

\begin{figure}
    \centering
    \subfloat[]{
        \includegraphics[width=0.5\textwidth]{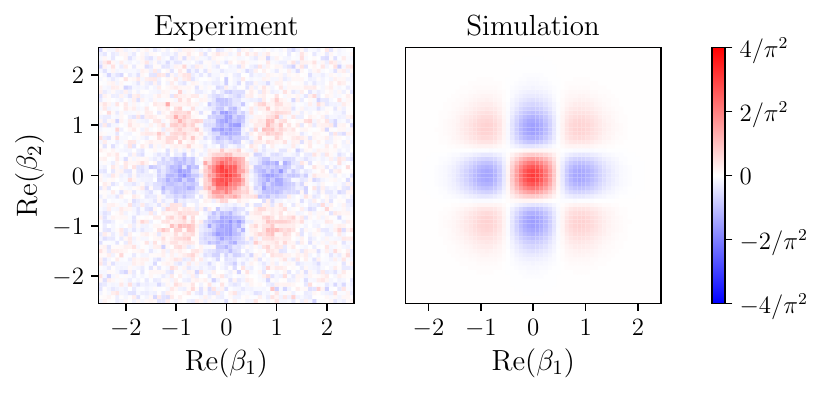}
    }
    \vfill
    \subfloat[]{
        \includegraphics[width=0.5\textwidth]{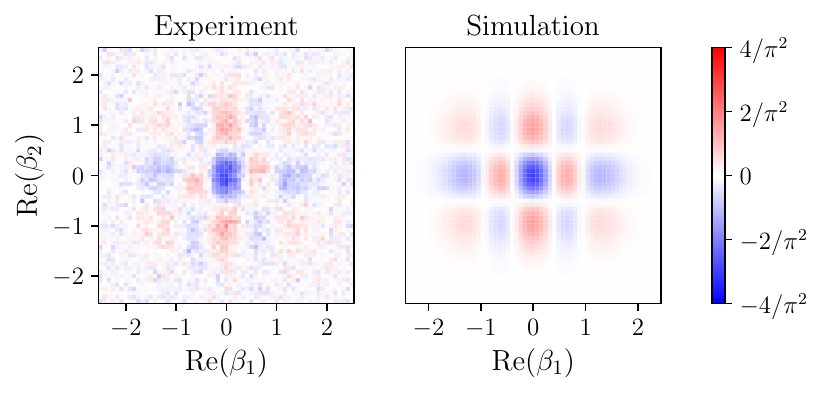}
    }
    \caption{\label{fig:wigner-fock} Wigner distributions $W(\beta_1, \beta_2)$ of two-mode Fock states. (a) Fock state $\ket{1}\ket{1}$; (b) Fock state $\ket{2}\ket{1}$. The ``Experiment'' and ``Simulation'' columns depict the experimentally measured and numerically simulated Wigner functions, respectively. See Appendix~\ref{sect:numerical-simulation} for the noise-aware simulation. The experiments are repeated by 300 shots per data point.}
\end{figure}

Figure~\ref{fig:wigner-fock} shows the two-dimensional quantum state tomograms of the two-mode Fock states. $\beta_1$ and $\beta_2$ denote the phase-space displacements applied to the modes $1$ and $2$, respectively, for sampling the Wigner function. The RMSEs between the experiment and noise-aware simulation are 0.026 and 0.028 for Fig.~\ref{fig:wigner-fock}(a) and (b), corresponding to 3.2\% and 3.5\% of the full Wigner-function value range $8 / \pi^2$, respectively. These values indicate good agreement and validate the correct implementation of displacement operations and the Wigner function measurement. Moreover, it clearly reveals negativity of the Wigner function, as expected for non-classical states such as Fock states or ECSs. Each two-dimensional plane consists of $61 \times  61 = \text{3,721}$ phase-space points, 300 experimental repetitions per point. The measurement at this single phase-space point takes approximately 4 seconds. Including interleaved calibrations of the Raman beam positions, the carrier Rabi frequency, the motional mode frequencies, the spin-dependent force amplitude balance, and the spin-dependent beam splitter offset detuning (see Appendix~\ref{sect:calibration-routine}), which account for 30--40\% of the total time, the entire data acquisition took about 320 minutes. The data set presented in Fig.~\ref{fig:wigner-fock} has a slight positive bias of about $1\%$ of the full scale of the data, which we believe originates from the error in detecting the $\ket{\downarrow}$ state.

\subsection{Tomography of Two-mode Entangled Coherent States \label{sect:tomography-ecs}}

\subsubsection{State Preparation}

To prepare an entangled coherent state of motion, we apply the spin-dependent forces separately to each mode. Although such a state can also be generated using a single detuned spin-dependent force applied simultaneously to both modes when they are nearly degenerate~\cite{Jeon2024Experimental}, we instead apply resonant spin-dependent forces to each mode individually, as the two motional modes are far apart in the current setup. By sequentially applying spin-dependent forces to the sideband-cooled initial state $\ket{\downarrow}\ket{0}\ket{0} = 1/\sqrt{2}(\ket{+}+\ket{-})\ket{0}\ket{0}$, with displacements $\alpha_1$ and $\alpha_2$ for the modes $1$ and $2$, respectively, the following spin--motion-entangled state is generated:
\begin{equation}\label{eq:ecs_wf}
    \begin{split}
    \ket{\psi_0}=\frac{1}{2}(\ket{\downarrow}(\ket{\alpha_1}\ket{\alpha_2}+\ket{-\alpha_1}\ket{-\alpha_2})+\ket{\uparrow}(\ket{\alpha_1}\ket{\alpha_2}-\ket{-\alpha_1}\ket{-\alpha_2}))\\
    =\frac{1}{2}(\mathcal N_+ \ket{\downarrow}\ket{\mathrm{Even}} + \mathcal N_- \ket{\uparrow}\ket{\mathrm{Odd}}),
    \end{split}
\end{equation}
where $\ket{\mathrm{Even(Odd)}}$ denotes the normalized even (odd) entangled coherent state (ECS), whose joint parity is even (odd), and $\mathcal N_\pm = \sqrt{2 \pm 2e^{-2(|\alpha_1|^2+|\alpha_2|^2)}}$ are the inverse of normalization factors: $\ket{\mathrm{Even (Odd)}} = \frac{1}{\mathcal{N}_\pm} \left( \ket{\alpha_1}\ket{\alpha_2} \pm \ket{-\alpha_1}\ket{-\alpha_2} \right)$. Therefore, by detecting the spin state after the sequential application of the spin-dependent forces, we probabilistically prepare one of the ECSs. However, if the $\ket{\uparrow}$ state, which corresponds to the bright state, is detected, the motional state is not preserved due to momentum recoil from photon scattering during fluorescence detection. To circumvent this effect when generating an odd ECS, we flip the spin state at the beginning of the sequence using a carrier $\pi$-rotation. The subsequent spin-dependent forces then map $\ket{\downarrow}$, the dark state, to $\ket{\mathrm{Odd}}$.

\subsubsection{Wigner Function Measurement \label{sect:ecs-wigner-measurement}}

\begin{figure*}
    \centering
    \subfloat[]{
        \includegraphics[height=0.28\textwidth]{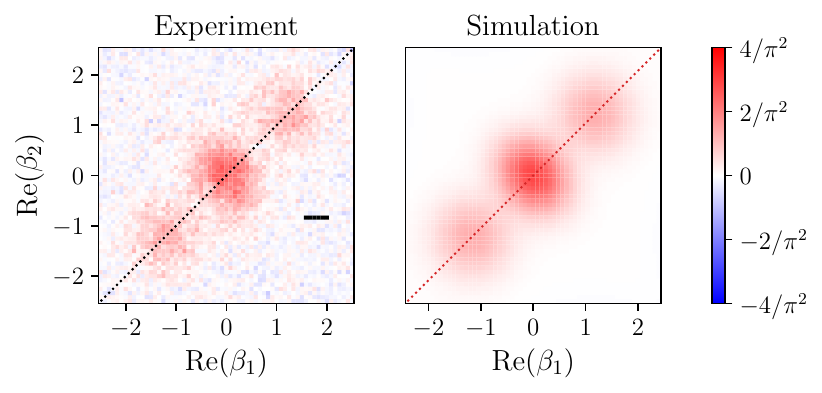}
    }
    \subfloat[]{
        \includegraphics[height=0.28\textwidth]{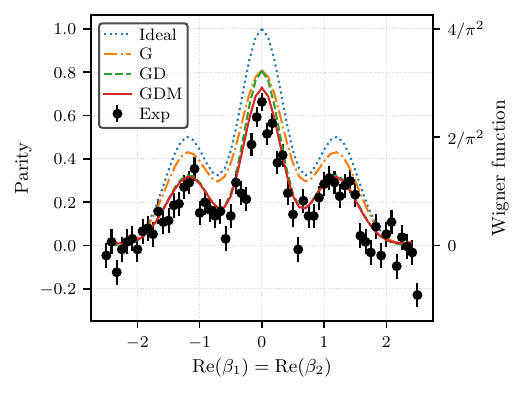}
    }
    \vfill
    \subfloat[]{
        \includegraphics[height=0.28\textwidth]{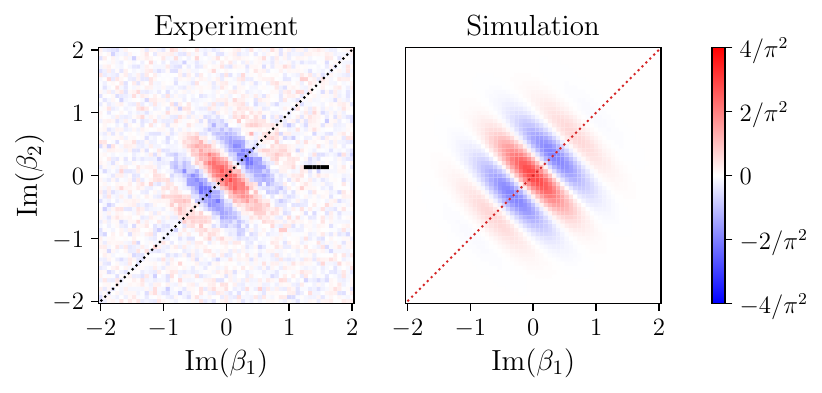}
    }
    \subfloat[]{
        \includegraphics[height=0.28\textwidth]{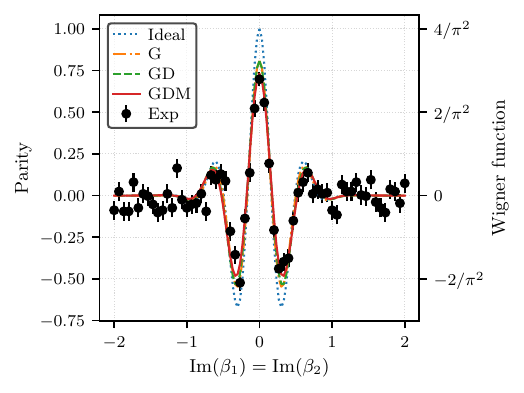}
    }
    \caption{\label{fig:wigner-ecs-even} Wigner distributions $W(\beta_1, \beta_2)$ of an even ECS $\ket{\mathrm{Even}}$ with $\alpha_1=\alpha_2=1.2$, shown as two-dimensional slices of the four-dimensional phase space: (a) real-real plane and (c) imaginary-imaginary plane. Each column represents the experimentally measured and numerically simulated Wigner distributions, respectively. Black regions indicate points for which data were not recorded due to an unintentional acquisition omission identified during post-processing. These points were not included in density matrix estimation. The 1D slices of the Wigner distributions along the dotted diagonal lines are shown in greater detail with noise-aware simulation results: (b) real-real diagonal and (d) imaginary-imaginary diagonal. \textbf{G}, \textbf{D}, and \textbf{M} in the legend denote the state generation, displacement, and measurement steps, respectively, during which the noise is applied in the simulation. The experiments are repeated by 300 shots per data point, and the error bars are derived from the quantum projection noise of the measured spin populations.}
\end{figure*}

\begin{figure*}
    \centering
    \subfloat[]{
        \includegraphics[height=0.28\textwidth]{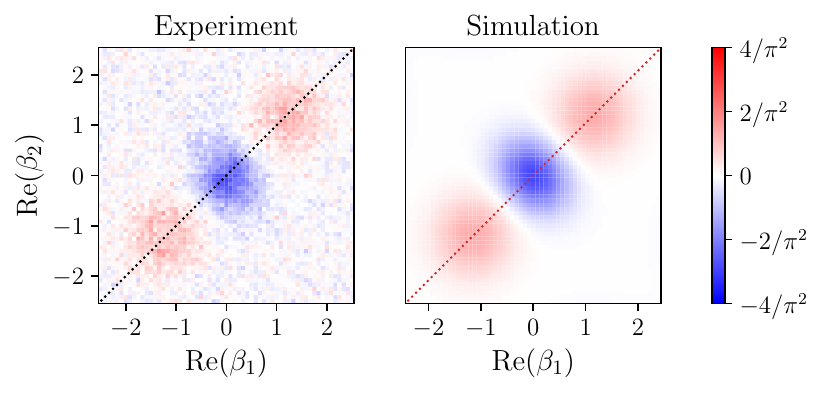}
    }
    \subfloat[]{
        \includegraphics[height=0.28\textwidth]{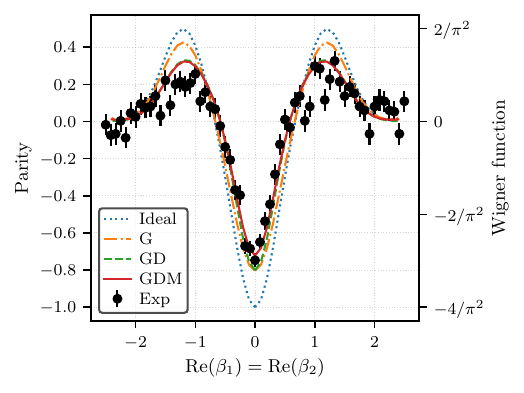}
    }
    \vfill
    \subfloat[]{
        \includegraphics[height=0.28\textwidth]{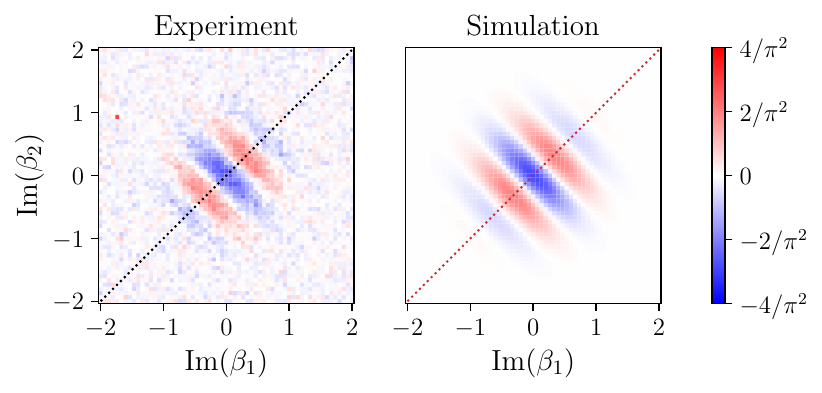}
    }
    \subfloat[]{
        \includegraphics[height=0.28\textwidth]{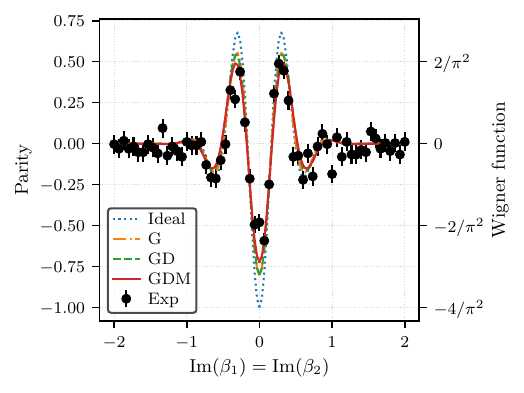}
    }
    \caption{\label{fig:wigner-ecs-odd} Wigner distributions $W(\beta_1, \beta_2)$ of an odd ECS $\ket{\mathrm{Odd}}$ with $\alpha_1=\alpha_2=1.2$, shown as two-dimensional slices of the four-dimensional phase space: (a) real-real plane and (c) imaginary-imaginary plane. The 1D slices of the Wigner distributions along the dotted diagonal lines are shown in greater detail with noise-aware simulation results: (b) real-real diagonal and (d) imaginary-imaginary diagonal. \textbf{G}, \textbf{D}, and \textbf{M} in the legend denote the state generation, displacement, and measurement steps, respectively, during which the noise is applied in the simulation. The experiments are repeated by 300 shots per data point, and the error bars are derived from the quantum projection noise of the measured spin populations.}
\end{figure*}

Figure~\ref{fig:wigner-ecs-even} shows the experimentally measured Wigner function of an even ECS with $\alpha_1=\alpha_2=1.2$. In panels (a) and (c), two-dimensional slices of the four-dimensional phase space distribution are obtained by displacing the quantum state along the real and imaginary axes of the phase spaces of mode $1$ and $2$. The RMSEs between the experimental results and the numerical simulations are 0.030 and 0.029 for the real-real and imaginary-imaginary planes, corresponding to 3.7\% and 3.6\% of the full Wigner-function value range $8 / \pi^2$, respectively, indicating good agreement. The experimentally measured Wigner distributions show all the expected features such as negative Wigner function values and interference fringes characteristic of nonclassical states.

Figure~\ref{fig:wigner-ecs-even}(b) and (d) illustrate the Wigner function along the one-dimensional diagonal dotted lines indicated in Fig.~\ref{fig:wigner-ecs-even}(a) and (c), respectively. \textbf{Ideal} refers to the theoretical Wigner function of a perfect ECS. The other curves represent the noise-aware numerical simulations that include motional heating and secular frequency fluctuations which we believe mainly originates from the trap potential stabilization circuit picking up the mains power oscillation in the lab (see Appendices~\ref{sect:error/heating-rate} and \ref{sect:error/finite-motional-coherence-time} for details on noise sources, and Appendix~\ref{sect:numerical-simulation} for the simulation method). We divide the experimental sequence into three steps: generation, displacement, and measurement, labeled as \textbf{G}, \textbf{D}, and \textbf{M} in Fig.~\ref{fig:wigner-ecs-even}. To isolate the contribution of each step to the observed deviation from the ideal Wigner function values, we incrementally incorporate each step into the simulation. \textbf{G} (generation) shows the Wigner functions of a state obtained from numerical simulation of the noisy generation sequence alone. In this step, the parity of the generated state is reduced from unity due to motional heating, which occurs primarily during the 250~$\mu$s-long  spin measurement required to collapse the spin--motion entangled state into the ECS. \textbf{GD} (generation and displacements) includes the simulation of the displacement operators---the spin-dependent forces---used to translate the Wigner function, while the parity is still calculated directly, without simulating the spin-dependent beam splitter. Due to fluctuations in the secular frequency, the direction of the displacement force varies, causing increased deviation of the estimated Wigner function from the ideal as the displacement amplitude increases. \textbf{GDM} (generation, displacements, and measurement) corresponds to the simulation of the entire Wigner function measurement sequence, including both the displacement operations and the joint parity measurement. 

The results for the odd ECS are also shown in Fig.~\ref{fig:wigner-ecs-odd}. We can clearly see that the Wigner distribution is now the negative mirror of the even ECS distribution as expected. The RMSEs between the experiment and simulation data are 0.029 and 0.028 for the real-real and imaginary-imaginary planes, corresponding to 3.6\% and 3.5\% of the full Wigner-function value range $8 / \pi^2$, respectively, indicating good agreement. These observations confirm that our numerical simulations, which account for all experimental imperfections, accurately reproduce the measured Wigner functions. Moreover, the generated state---whose Wigner function is shown as \textbf{G} in Fig.~\ref{fig:wigner-ecs-even}---is expected to violate the CHSH inequality, as discussed in Appendix~\ref{sect:chsh-inequality-violation}.

Lastly, Table~\ref{tbl:error_budget} summarizes how each step of the experimental sequence and the error source present in the system contributes to the reduction of Wigner function contrast for the even ECS in the imaginary-imaginary plane. We assume that the off-resonant carrier excitation can be ignored by choosing proper interaction times as described in Appendix~\ref{sect:error/off-resonant-carrier}. The largest contrast loss comes from the quantum state generation step \textbf{G}, with almost equal contributions from non-zero initial temperature and motional heating. During the displacement step \textbf{D}, motional dephasing---caused by secular frequency fluctuations---is the dominant source of contrast loss, as it perturbs the direction of the spin-dependent force. In the measurement step \textbf{M}, motional heating again causes a large reduction in contrast. Overall, the majority of decoherence takes place in the generation and measurement step. These steps take a much longer time than the displacement step, because they involve the detection of the spin state and the $\pi$ beam-splitter pulse which is used to map phonon parity to the spin state. Therefore, the ion is subject to more motional heating in these steps, resulting in higher contrast loss.

\begin{table}[h]
\centering
\begin{tabular}{cccc}
\toprule
\textbf{Step} & \textbf{Contrast Loss} & \textbf{Error Source} & \textbf{Contrast Loss by Source} \\
\midrule
\multirow{3}{*}{\shortstack{Generation (\textbf{G}) \\ $350\,\mu\mathrm{s}$}} & \multirow{3}{*}{$17.2\%$} & Initial temperature ($\overline{n}_1,\overline{n}_2 = 0.03$) & $8.35\%$ \\
                            &                    & Heating & $8.82\%$ \\
                            &                    & Motional dephasing & $0.01\%$ \\
\midrule
\multirow{2}{*}{\shortstack{Displacement (\textbf{D}) \\ $0\text{--}200\,\mu\mathrm{s}$}} & \multirow{2}{*}{$0.69\%$} & Heating & $0.11\%$ \\
                            &                    & Motional dephasing & $0.58\%$ \\
\midrule
\multirow{2}{*}{\shortstack{Measurement (\textbf{M}) \\ $550\,\mu\mathrm{s}$}} & \multirow{2}{*}{$8.15\%$} & Heating & $8.15\%$ \\
                            &                    & Motional dephasing & $0.0\%$ \\
\midrule
\multirow{1}{*}{Total} & \multirow{1}{*} & & $26.02\%$ \\
\bottomrule
\end{tabular}
\caption{\label{tbl:error_budget} Breakdown of numerically simulated Wigner function contrast loss by step and error source in the even ECS imaginary-imaginary plane. The approximate duration of each step is indicated next to its name. The duration of step \textbf{D} is proportional to the displacement amplitudes. Longer steps (\textbf{G} and \textbf{M}) cause larger loss in contrast mainly due to heating. Heating rates of $\dot {\overline n}_1=6.1$ and $\dot {\overline n}_2=39.0\,\mathrm{quanta/s}$ are used, along with a 60-Hz mode frequency modulation amplitude of $\Delta=150\,\mathrm{Hz}$ to model motional dephasing. In step \textbf{M}, motional dephasing is not simulated because the mode frequency noise is expected to be in-phase for both modes, which makes it effectively invisible to the spin-dependent beam splitter interaction. See Appendices~\ref{sect:error/heating-rate} and \ref{sect:error/finite-motional-coherence-time} for details.}
\end{table}

\subsubsection{Model-Based Density Matrix Estimation \label{sect:ecs-density-matrix}}

With the QST of the even and odd ECSs in hand, we estimate the density matrix of the experimentally generated state, $\rho_\mathrm{est}$, in the two-mode Fock basis. Since the measured data consist of two-dimensional slices, we use a constrained model to infer the density matrix that best explains the observed data rather than attempting a full, unconstrained reconstruction. Our model takes into account the major error sources identified by numerical simulation as in Table~\ref{tbl:error_budget}: initial temperature of each mode, motional heating during the generation and measurement phase, and motional dephasing in the displacement phase. In the estimation process, we fit a model Wigner distribution $W(\beta_1, \beta_2;\alpha_1, \alpha_2, \overline{n}_1, \overline{n}_2, \gamma_1, \gamma_2, \delta_1, \delta_2, c)$ to the two-dimensional slices shown in Figs.~\ref{fig:wigner-ecs-even} and \ref{fig:wigner-ecs-odd}. $\beta_1$ and $\beta_2$ are the displacements for each mode in their respective phase space. The fitted parameters are $\alpha_1, \alpha_2, \overline{n}_1, \overline{n}_2, \gamma_1, \gamma_2, \delta_1, \delta_2$ and $c$: $\alpha_{i}$, $\overline{n}_i, \gamma_{i}$ and $\delta_{i}$ are the amplitude of the cat state, initial phonon number, heating rate and motional dephasing rate for $i$-th mode. $c$ is the error in the $\ket{\downarrow}$ state detection. In the respective density matrix estimation of the even and odd ECSs, both the real-real plane slice and imaginary-imaginary slice are used simultaneously. We note that this estimation is based on the assumption that our error model is correct. In the following analysis, the fidelity of the generated state is defined as $F=(\operatorname{Tr}(\sqrt{\sqrt{\rho_0}\rho_\mathrm{est}\sqrt{\rho_0}}))^2$, where $\rho_0$ is the density matrix of the ideal target state. $\rho_\mathrm{est}$ incorporates the errors in the measurement process, and the fidelity and purity values include imperfections of the measurement sequence which includes the SBS operation and the spin state measurement.

\begin{figure}
    \centering
    \subfloat[]{
        \includegraphics[width=0.45\textwidth]{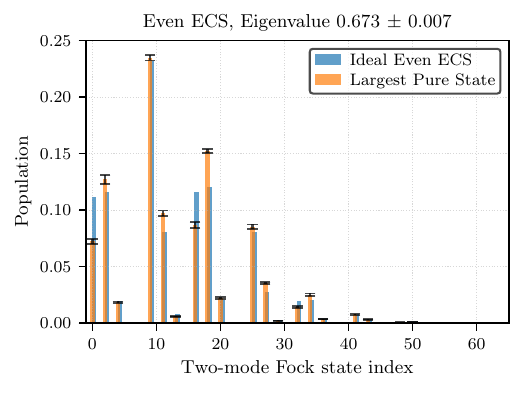}
    }
    \hfill
    \subfloat[]{
        \includegraphics[width=0.45\textwidth]{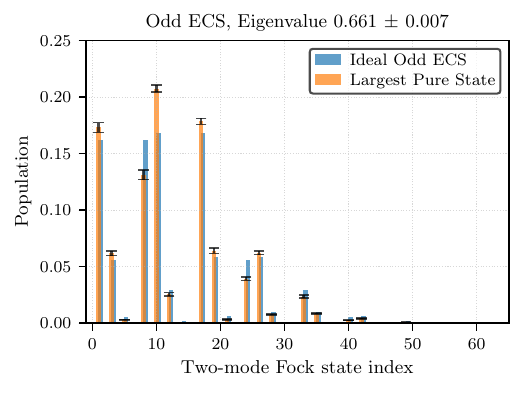}
    }
    \caption{\label{fig:wigner_dm} Most dominant eigenstates in the estimated density matrices. (a) The eigenvector of $\rho_\mathrm{est}$ of the even ECS with the largest eigenvalue is compared with the population distribution of the ideal even ECS. The horizontal axis is the index $i$ for the two mode Fock state, which is defined as $i= n_1 (n_\mathrm{max}+1)+n_2$ for $\ket{n_1}\ket{n_2}$ where $n_1$ and $n_2$ are the phonon numbers of each mode and $n_\mathrm{max}=7$ is the phonon cutoff of the density matrix. This eigenvector composes $67.3(7)\%$ of the estimated density matrix. (b) The two-mode Fock state population of the most dominant eigenvector of the odd ECS. This state makes up $66.2(8)\%$ of the density matrix. Error bars were calculated from statistical bootstrapping using 50 samples.
    }
\end{figure}

We consider phonon numbers from $0$ to $7$ for each mode. With a phonon number cut-off of $n_\mathrm{max}=7$, more than $99.9\%$ of the population in the target state is accounted for. For the even ECS, the fitted density matrix yields a fidelity of $0.666(8)$ with the target state and purity of $0.474(10)$. In the case of the odd ECS, the fidelity with respect to the target state is $0.655(8)$, and the purity is $0.457(9)$. The rest of the fitted parameters are compiled in Appendix~\ref{sect:model-based-density-matrix-estimation}. For the even ECS, The largest eigenvalue is $0.673(7)$ and the corresponding eigenvector is very similar to the target state which is the even ECS as shown in Fig.~\ref{fig:wigner_dm}(a). The largest eigenvalue of the odd ECS density matrix is $0.661(7)$, and its population distribution matches that of the pure odd ECS closely as compared in Fig.~\ref{fig:wigner_dm}(b). The standard deviations for the fidelity, purity and eigenvalues were estimated by statistical bootstrapping in which we generated 50 synthetic datasets by resampling from the experimentally measured binomial distribution and fitted their respective density matrices. Also, the two datasets exhibit error in $\ket{\downarrow}$ state measurement which affects the contrast and offset in Wigner function measurement. The inferred $\ket{\downarrow}$ state measurement error is $c=0.0090(9)$ and $0.0129(8)$ for the original even and odd datasets, respectively. The quality of these fits can be quantified using the reduced chi-squared values, which are respectively $1.241$ and $1.338$. It indicates a good match between the model and the observed data.

Also, we check if the state we generated is consistent with the Peres--Horodecki criterion~\cite{Horodecki1996, Peres1996} for entanglement. The criterion states that for a bipartite density matrix, if its partial-transposed form contains a negative eigenvalue, then the two systems are entangled. We partial-transpose $\rho_\mathrm{est}$ to get $\rho_\mathrm{est}^{\Gamma_{1}}$ in which the indices of mode $1$ are flipped. We then calculate the eigenvalues of $\rho_\mathrm{est}^{\Gamma_{1}}$. We find that the matrices for both the even and odd ECSs have a few negative eigenvalues. For the even ECS, the most negative eigenvalue is $-0.307(4)$, and for the odd ECS, it is $-0.306(6)$. This indicates that these states are entangled under the Peres--Horodecki criterion. We note that this is a model-dependent analysis and not a certification of entanglement. 

\subsection{Parity-Filtered Recovery of a Heated Fock State}\label{sect:recovery-parity-selection}

Motional heating is one of the dominant decoherence mechanisms in trapped-ion continuous-variable systems that use phonons: each heating event adds a phonon to one of the modes and therefore flips the joint parity. A single-shot joint-parity measurement can thus act as a herald for those heating errors. By retaining only the outcomes with the same parity as the starting parity, we effectively filter out corrupted trajectories and recover the original bosonic states in post-selection--based error mitigation.

\begin{figure}[t]
  \centering
  \begin{minipage}[b]{0.49\linewidth}
    \subfloat[]{%
      \includegraphics{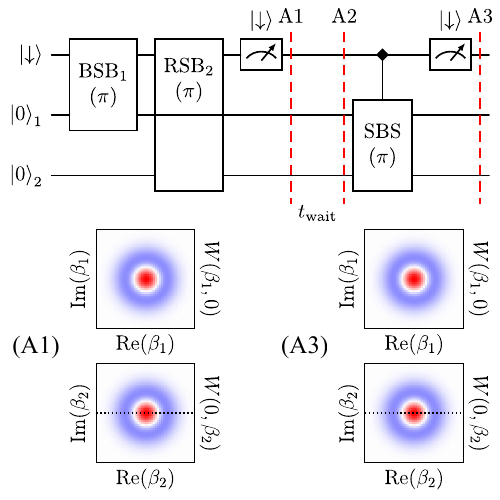}}
  \end{minipage}
  \begin{minipage}[b]{0.49\linewidth}
    \subfloat[]{%
      \includegraphics[width=0.7\linewidth]{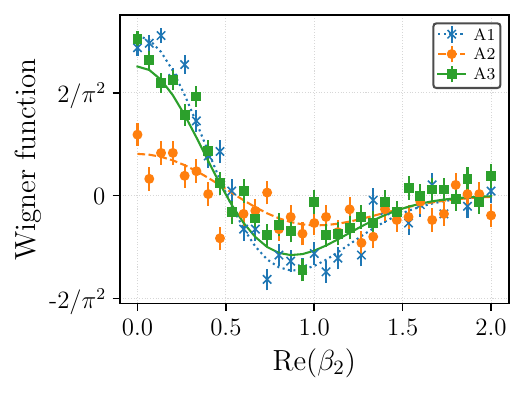}}
    \vfill
    \subfloat[]{%
      \includegraphics[width=0.6\linewidth]{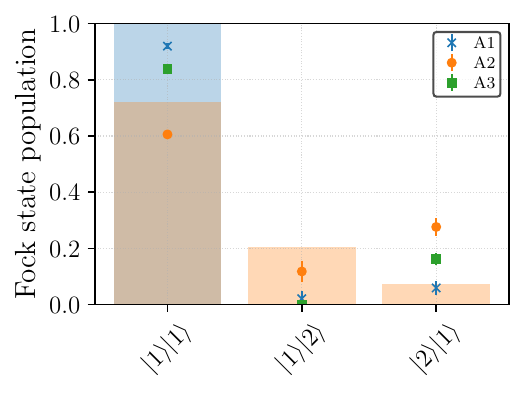}}
  \end{minipage}
  \caption{Recovery of a two-mode Fock state after heating via parity filtering.
    (a)~Pulse sequence used to prepare the Fock state $\ket{1}_1\ket{1}_2$, allow motional thermalization, and perform a filtering based on joint-parity measurement. The plots below the circuit illustrate the expected Wigner function of the quantum states at \textbf{A1} and \textbf{A3}. Wigner functions in (b) are measured along the dotted lines in these plots.
    (b)~Measured one-dimensional Wigner slices $W(0, \beta_2)$, where $\text{Im}(\beta_2)=0$, taken at three instances labeled \textbf{A1}--\textbf{A3} in (a): \textbf{A1} (blue crosses) corresponds to the initial state after preparation, \textbf{A2} (orange circles) is after 10-ms thermalization, and \textbf{A3} (green squares) is after parity-based post-selection. Error bars represent spin-projection noise, computed from 300 repetitions per data point. Curves represent the numerical simulation results, acquired by fitting the Fock state populations.
    (c)~Fitted Fock-state populations for \textbf{A1}--\textbf{A3}. Blue and orange bars show the theoretical populations for the ideal state $\ket{1}_1\ket{1}_2$ (\textbf{A1}) and after 10~ms of heating (\textbf{A2}), respectively. Error bars are estimated using 1,000 bootstrap resamples.}
  \label{fig:parity-filtering}
\end{figure}

We test this error-mitigation protocol with the two-mode Fock state $\ket{1}_1\ket{1}_2$. Figure~\ref{fig:parity-filtering}(a) shows the full pulse sequence where we first prepare the two‑mode Fock state by sequentially applying blue- and red-sideband $\pi$-pulses, followed by spin-state filtering to remove any population in $\ket{\uparrow}$, identified by fluorescence counts exceeding a predefined threshold.
We then wait for $\smash{t_{\text{wait}} = 10\,\text{ms}}$ to allow motional thermalization. After the wait, we perform a joint‑parity measurement using $\operatorname{SBS}(\pi)$ and post‑select on the $\ket{\downarrow}$ outcome to filter out odd‑parity events. To visualize the recovery of the original state, we measure one-dimensional Wigner slices $W(0,\beta_2)$ at three instances labeled \textbf{A1} (immediately after preparation), \textbf{A2} (after the 10-ms wait), and \textbf{A3} (after post-selection).
The joint-parity measurement steps are not shown in the quantum circuit in Fig.~\ref{fig:parity-filtering}(a) to avoid confusion, and the results are displayed in Fig.~\ref{fig:parity-filtering}(b).
The data taken at \textbf{A2}, after 10-ms thermalization, are shown as orange circles and differ significantly from those at \textbf{A1}, taken immediately after the state generation and shown as blue crosses.
The Wigner distribution measured at \textbf{A3} are shown as green squares. After post-selection, the Wigner distribution closely overlap with that at \textbf{A1}, demonstrating that the initial Fock state is restored when odd-parity shots are filtered out.
To quantify how well the state recovers, we estimate the populations of $\ket{1}_1\ket{1}_2$, $\ket{1}_1\ket{2}_2$, and $\ket{2}_1\ket{1}_2$ by fitting the experimental data to a basis of Wigner distributions obtained by simulating the Wigner-function measurement for each ideal Fock state. The fit coefficients give the corresponding Fock-state populations. It can be seen that the population of $\ket{1}_1\ket{1}_2$ which has decreased to $0.6$, was recovered to above $0.8$ after post-selection. The error bars of the populations are estimated via bootstrap with 1,000 resamples.
These results confirm that the single-shot measurement of joint-parity can serve as an effective error-mitigation tool against motional heating in trapped-ion continuous-variable quantum states.

\section{Conclusions and Outlook}
In this study, we have realized a single-shot measurement of the joint parity of the phonons in the motional modes of a trapped ion. This was implemented by driving a bichromatic spin-dependent beam splitter transition that swaps the quantum states of the two participating motional modes. The operation encodes a phonon-number--dependent phase in the spin state, which can be read out using the standard fluorescence measurement technique. This operation allowed us to perform quantum state tomography on various continuous-variable quantum states, such as two-mode Fock states and entangled coherent states. Since it was not guaranteed to be informationally complete, the experimentally measured tomogram was used to estimate the density matrix of the system using a model that takes various error sources into account. The density matrix was then used to calculate various quantum informational quantities such as the state fidelity, purity, eigenvalues and eigenvectors. Lastly, using the single-shot nature of the measurement, we were able to detect parity flip in two-mode Fock states. By post-selecting the experimental runs in which the parity remained unchanged, we were able to partially recover the original quantum state, even after it had undergone significant decoherence. 

Our work enables two key capabilities: (i) direct and efficient quantum state tomography via Wigner function measurement with minimal experimental overhead, and (ii) single-shot detection of joint-parity flip in a multi-mode system, a crucial element in continuous-variable quantum error correction~\cite{fluhmann_encoding_2019, wang_schrodinger_2016, Guillaud2019, Putterman2025Hardware, de_neeve_error_2022}. Notably, the single-shot measurement of phonon parity enables syndrome extraction which detects single phonon loss or gain errors in logical encodings such as the cat code and the binomial code~\cite{Michael2016, Guillaud2019, Ni2023, Putterman2025Hardware} with a trapped ion. The interaction is also relevant to the field of quantum metrology, as the parity operator serves as an optimal observable in interferometric measurements, capable of saturating the Cramér-Rao bound \cite{Gerry2010parity, Joo2011}. Therefore, this study represents an important addition to the toolbox of operations available for measuring and controlling continuous-variable quantum states in the platform.

The current measurement scheme perturbs the motional state of the ion if we detect a bright state. However, this problem can be circumvented when working with an ion chain that consists of at least three ions. We can transfer the parity information using a controlled-NOT (CNOT) gate to the spin of another ion which does not participate in the normal modes of interest. A similar concept was demonstrated in Ref.~\cite{hou_coherent_2024}. As an example, this scheme can be realized by using the two radial out-of-phase modes of a three-ion chain as the carrier of quantum information, and using one of the radial zig-zag modes or axial modes for the CNOT  gate. When we selectively detect the center ion which does not participate in the out-of-phase motion, the photon recoil from detecting a bright state does not destroy the motional modes.

Finally, we note that the multi-mode parity measurement is scalable. The spin-dependent beam splitter operations for different pairs of motional modes can be applied back to back simply by changing the detunings of the bichromatic Raman transition. This means that the phonon number parity of $2N$ modes can be measured by $N$ operations. Although mode crowding and off-resonant driving of neighboring modes will be an issue as the number of modes are increased, this can be mitigated by reducing the speed of interaction appropriately~\cite{Chen2023Scalable}.

\begin{acknowledgments}
We thank H. Kwon and Y. Teo for helpful discussions. This work was supported by the Institute for Information \& Communications Technology Planning \& Evaluation (IITP) grant (No. RS-2022-II221040), and the National Research Foundation of Korea (NRF) grant (No. RS-2020-NR049232, No. RS-2024-00442855, No. RS-2024-00413957), all of which are funded by the Korean government (MSIT).
\end{acknowledgments}

\section{Author Contributions}
H. Jeon conceived the idea, conducted theoretical calculations and proof-of-concept experiments, and led the estimation of the density matrix and its analysis. J. Kang optimized the spin-dependent beam splitter interaction, acquired the experimental data, and performed noise-aware numerical simulations for data analysis. W. Choi, K. Kim, and J. You helped maintain the experimental setup and contributed to research discussions. J. Kang and J. You also contributed to improvements in the experimental control software system. T. Kim supervised the project, including research planning, experimental design, data interpretation, and manuscript preparation.

\appendix

\section{Model-Based Estimation and Analysis of the Density Matrix} \label{sect:model-based-density-matrix-estimation}
Due to the limited data set, we estimated the density matrices of the experimentally generated states by assuming a model that contains several free parameters and fitting it to the data rather than attempting a full reconstruction. The model considers the residual phonon population in each mode after sideband cooling $(\overline{n}_{1,2})$, motional heating rates in each mode $(\gamma_{1,2})$, motional dephasing rates of each mode $(\delta_{1,2})$, the size of the cat state in each mode $(\alpha_{1,2})$ and the $\ket{{\downarrow}}$ state detection error $(c)$, which is equivalent to $q_\downarrow$ in Appendix~\ref{sect:error/qubit-detection}.

The model first generates a mixed state density matrix given initial mean phonon numbers $\overline{n}_1$ and $\overline{n}_2$. It then applies heating for the duration of the state generation phase, which is $350\ \mu\mathrm{s}$. Most of this duration is spent on detecting the spin state to generate the ECS and the displacement pulses on each mode is only a few tens of microseconds. Therefore, we consider heating to be the dominant dephasing mechanism. In the displacement phase, a ``rotational smearing'' is applied to simulate the smearing of displacement angle in the phase space of each mode, that originates from the fluctuation in the trap frequency or motional dephasing. As shown in Fig.~\ref{fig:wigner-ecs-even} and Fig.~\ref{fig:wigner-ecs-odd}, because of the long duration of the displacement pulse, the effect of motional dephasing noticeably affects the measured Wigner function. Since the Wigner transform on the density matrix is linear and a rotation operation on the density matrix directly transfers to the Wigner function in the phase space, we first construct a mixed density matrix in the following form:

\begin{equation}
    \overline{\rho}=\int_{-4\sigma_1}^{4\sigma_1}\int_{-4\sigma_2}^{4\sigma_2}d\phi_1\,d\phi_2 \mathcal{N}(\phi_1; 0, \sigma_1^2)\mathcal{N}(\phi_2; 0, \sigma_2^2)R^\dagger_1(\phi_1)R^\dagger_2(\phi_2)\rho_{\textbf{G}}\,R_1(\phi_1)R_2(\phi_2),
\end{equation}
where $\mathcal{N}(\phi_j;0,\sigma_j^2)$ is a normal distribution for a random variable $\phi_j$ centered at $0$ with a variance of $\sigma_j^2$ and $R_j(\phi_j)=\exp \left(i \phi_j a_j^\dagger a_j \right)$ is a rotation operator acting on the $j$-th mode. $\rho_{\textbf{G}}$ is the density matrix of the system after the generation phase. The integral is approximated by sampling $20$ evenly spaced values for $\phi_1$ and $\phi_2$ respectively. The width of smearing is related to the dephasing rate as $\sigma_j=\sqrt{\delta_j\,T_\mathrm{eff}}$ where $T_\mathrm{eff}$ is the effective average duration of the displacement pulses used in the tomography. $\delta_j$ is the dephasing rate for the $j$-th mode which is defined as $\delta_j=2/\tau_{01,j}$ where $\tau_{01,j}$ is the exponential decay constant of the motional coherence time of the $j$-th mode for a superposition of $0$ and $1$ phonon states. Points in the phase space farther away from the origin are more strongly affected by rotational smearing, with the strength of the effect increasing proportionally to $r_k^2$, where $r_k$ denotes the distance of the $k$-th point from the origin. This is because when the Wigner function in the polar coordinates, $W(r,\theta + \delta\theta)$, is Taylor expanded in $\delta\theta$, the second-order term in the Taylor expansion in the angular direction is the leading order term. We can calculate the effective average duration of the displacement pulses as $T_\mathrm{eff}=\Sigma_k r_k^2t_k/\Sigma_k r_k^2=\Sigma_kt_k^3/\Sigma_kt^2_k$ which is $75\ \mu\mathrm{s}$. $\overline{\rho}$ is then again subject to motional heating in the measurement phase for $550\ \mu\mathrm{s}$. In this phase, we do not simulate motional dephasing as the beam splitter operation has a much longer coherence time than the displacement operation. The final density matrix is converted to Wigner functions in the real-real and imaginary-imaginary planes, and they are compared with the experimental data using least-squares.

The results of the fit are presented in Table~\ref{tbl:fitting_results}. The fitted parameters are roughly in the same range as the independently measured values. The errors are calculated from the covariance matrix. There are small disagreements between the fitted values of the heating rates and dephasing rates and what is expected from independent experiments which we believe is due to the slow drift of these values in our setup. We note that these values fluctuate even within the same day. The origin of the change is unknown. The size of the cat, $\alpha_{1,2}$, shows a small deviation which indicates that the calibration of the amplitude of the displacement pulses has a small systematic error.

\begin{table}[h]
\centering
\setlength{\tabcolsep}{10pt} 
\renewcommand{\arraystretch}{1.3} 
\begin{tabular}{cccc}
\hline\hline
Parameter & Even ECS (Fitted) & Odd ECS (Fitted) & Expected \\
\hline
$n_1, n_2$        & $0.03(2),\;0.05(1)$ & $0.07(2),\;0.05(1)$ & $0.03,\;0.03$ \\
$\alpha_1, \alpha_2$ & $1.31(1),\;1.26(1)$ & $1.29(1),\;1.19(1)$ & $1.2,\;1.2$ \\
$\gamma_1, \gamma_2\ (s^{-1})$ & $11(6),\;52(7)$ & $28(8),\;20(9)$   &5.4--6.1, 16.5--39.0\\
$\delta_1, \delta_2\ (s^{-1})$  & $1{,}307(129),\;378(90)$ & $1{,}484(152),\;486(119)$ & 500--1,000, 500--1,000 \\
$c$               & $0.0090(9)$               & $0.0129(8)$          & $0.0$ \\
\hline\hline
\end{tabular}
\caption{\label{tbl:fitting_results} Comparison of fitted parameters for the even and odd entangled coherent states (ECSs). The standard errors of the fitted parameters are calculated from the covariance matrix.}
\end{table}

\section{CHSH Inequality Violation \label{sect:chsh-inequality-violation}}

\begin{figure}
    \centering
    \subfloat[]{
        \includegraphics[width=0.35\linewidth]{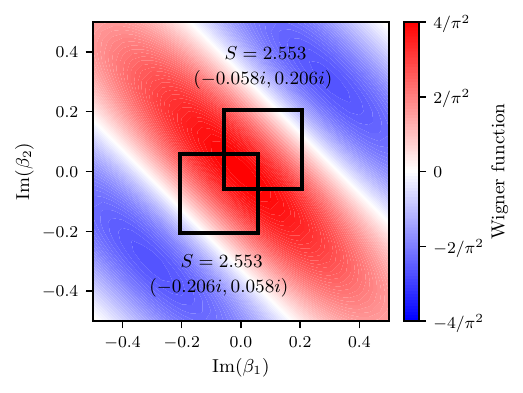}
    }
    \subfloat[]{
        \includegraphics[width=0.35\linewidth]{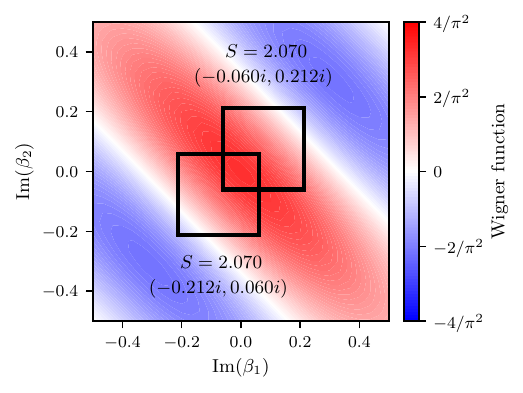}
    }
    \vfill
    \subfloat[]{
        \includegraphics[width=0.35\linewidth]{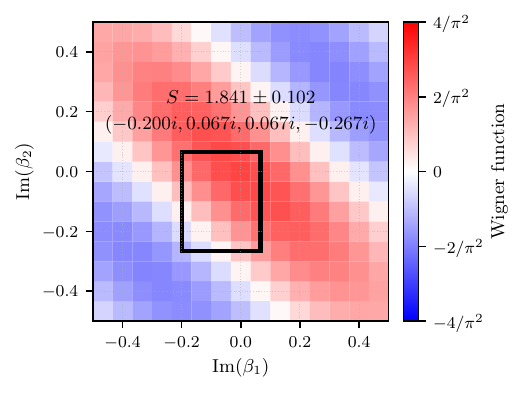}
    }
    \subfloat[]{
        \includegraphics[width=0.35\linewidth]{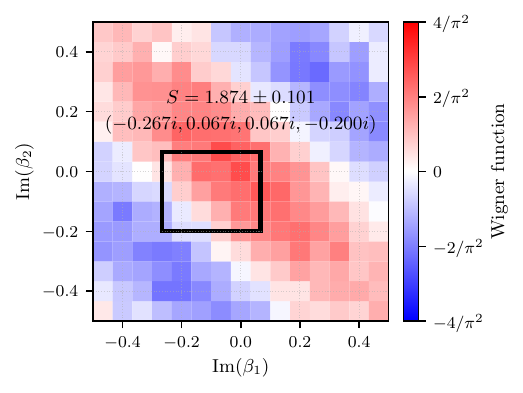}
    }
    \caption{\label{fig:chsh-even} Violation of the CHSH inequality using an ECS $\ket{\mathrm{Even}}$ with $\alpha_1=\alpha_2=1.2$: (a) ideal state and (b) simulated noisy state, corresponding to the \textbf{Ideal} and \textbf{G} curves in Fig.~\ref{fig:wigner-ecs-even}, respectively. In each image, the symmetrically positioned squares indicate the four displacement parameter pairs $\left(\beta_1^{(k)}, \beta_2^{(k')}\right)$, where $k,k' \in \{1, 2\}$, that maximize the CHSH parameter $S$. Since these pairs satisfy the symmetry relation $\beta_1^{(1)}=\beta_2^{(2)}$ and $\beta_1^{(2)}=\beta_2^{(1)}$, only the pair $\left(\beta_1^{(1)}, \beta_1^{(2)}\right)$ is labeled for each square. The background color maps represent the Wigner function $W(\beta_1, \beta_2)$.
    (c) and (d) show the maximum $S$ values for the end-to-end noise-aware simulation (\textbf{GDM} in Fig.~\ref{fig:wigner-ecs-even}) and experimental data shown in Fig.~\ref{fig:wigner-ecs-even}(c), respectively. Although they do not violate the inequality, they show great agreement. All the four displacements $\left(\beta_1^{(1)}, \beta_1^{(2)}, \beta_2^{(1)}, \beta_2^{(2)}\right)$ are shown for the panels (c) and (d) due to the discrete sampled points, which do not include the exact optimum points. The errors reflect the quantum projection noise.}
\end{figure}

The Clauser--Horne--Shimony--Holt (CHSH) inequality provides a testable criterion to distinguish quantum correlations from those explainable by local hidden variable theories~\cite{Clauser1969}. Using the Wigner function measurement scheme (or similarly with the characteristic function), it is possible to test this inequality with a two-mode ECS~\cite{wang_schrodinger_2016, Millican2025}. We take the single-mode parity operators $\mathcal{P}_1$ and $\mathcal{P}_2$ as observables, and use the displacement parameters $\beta_1$ and $\beta_2$ to define the measurement settings. The displaced joint parity operator is defined as
\begin{equation}
    \mathcal{P}(\beta_1, \beta_2) = D_1 (\beta_1) D_2 (\beta_2) \mathcal{P}_1 \mathcal{P}_2 D_1 (-\beta_1) D_2 (-\beta_2),
\end{equation}
whose expectation value is proportional to the two-mode Wigner function:
\begin{equation}
    \expval{\mathcal{P}(\beta_1, \beta_2)} = \frac{\pi^2}{4} W(\beta_1, \beta_2).
\end{equation}
Since the eigenvalues of the parity operators are $\pm 1$, this measurement setting is suitable for testing the CHSH inequality. We define the CHSH parameter as
\begin{equation}
    S = \expval{\mathcal{P}(\beta_1^{(1)}, \beta_2^{(1)})} - \expval{\mathcal{P}(\beta_1^{(1)}, \beta_2^{(2)})} + \expval{\mathcal{P}(\beta_1^{(2)}, \beta_2^{(1)})} + \expval{\mathcal{P}(\beta_1^{(2)}, \beta_2^{(2)})},
\end{equation}
where $\beta_j^{(k)}$ are the chosen displacement amplitudes in mode~$j$. As shown in Fig.~\ref{fig:chsh-even}(a), the calculated value of $S$ exceeds 2 for the ideal even ECS with $\alpha_1 = \alpha_2 = 1.2$, indicating the presence of quantum correlations. However, observing this violation experimentally is challenging in the current setup due to motional heating and mode frequency fluctuations, which reduce the fidelity of the joint parity measurement, as shown in Fig.~\ref{fig:chsh-even}(d) and expected by the end-to-end noise-aware simulation in Fig.~\ref{fig:chsh-even}(c). Nonetheless, when these experimental noise sources are included in the simulation, the generated state before displacement and parity measurement is still expected to violate the CHSH inequality, as shown in Fig.~\ref{fig:chsh-even}(b). 

\begin{figure}[t]
  \centering
  \subfloat[]{%
    \includegraphics[width=0.5\linewidth]{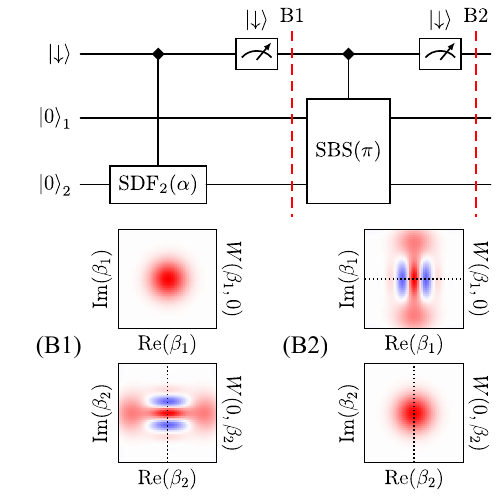}}
  \hfill
  \subfloat[]{%
    \includegraphics[width=0.45\linewidth]{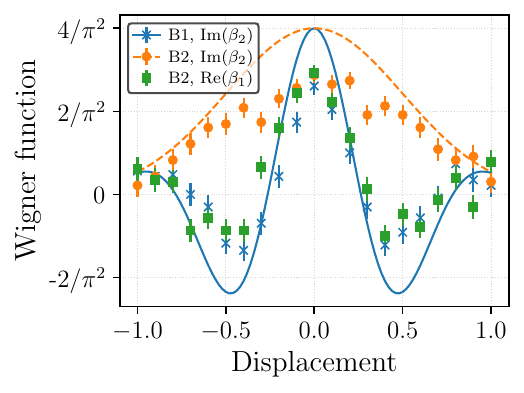}}
  \caption{Verification of the phonon-number--dependent phase and rotation using a cat state.
  (a)~Pulse sequence to prepare the state $\ket{0}_1(\ket{\alpha}_2 + \ket{-\alpha}_2)$ with $\alpha = 1.5$ using SDF, followed by heralding on $\ket{\downarrow}$. The state is then subjected to $\operatorname{SBS}(\pi)$, which transfers the cat state from mode~$2$ to mode~$1$ and rotates it in phase space by $\pi/2$. The plots below the circuit illustrate the expected Wigner functions of the quantum states at \textbf{B1} and \textbf{B2}. The dotted lines in these plots are the lines on which the Wigner functions in (b) were measured.
  (b)~One-dimensional Wigner slices measured at \textbf{B1} (before SBS) and \textbf{B2} (after SBS). The green squares at \textbf{B2} (mode~$1$) are plotted along $\rm{Re}(\beta_1)$ with $\rm{Im}(\beta_1) = 0$, while the blue crosses at \textbf{B1} (mode~$2$) are plotted along $\rm{Im}(\beta_2)$ with $\rm{Re}(\beta_2) = 0$. Their overlap confirms that the cat state is rotated by $\pi/2$ while the state is transferred from mode~$2$ to mode~$1$. The blue solid curve and orange dashed curve show theoretical Wigner slices for the ideal cat state and vacuum, respectively. Error bars denote spin-projection noise, computed from 200 repetitions per data point.}
  \label{fig:parity-select-cat}
\end{figure}

\section{Verification of the Phonon‑Number--Dependent Phase with a Cat State}
Equation~(\ref{eq:parity-sbs}) predicts a phonon‑number--dependent phase acquired during the $\operatorname{SBS}(\pi)$ pulse when the input motional state is composed of multiple Fock components. Also, the rotation of the motional state after the parity measurement can be calculated using Eq.~(\ref{eq:motion-phase-rotation}). To verify this prediction, we prepare the state $\ket{0}_1\bigl(\ket{\alpha}_2 + \ket{-\alpha}_2\bigr)$ with $\alpha = 1.5$ using a spin‑dependent force $\operatorname{SDF}_2(\alpha)$, followed by heralding on $\ket{\downarrow}$. We then apply $\operatorname{SBS}(\pi)$, which swaps the motional modes and rotates the cat state in phase space by $\pi/2$, yielding $\bigl(\ket{i\alpha}_1 + \ket{-i\alpha}_1\bigr)\ket{0}_2$.

Figure~\ref{fig:parity-select-cat}(b) shows one‑dimensional Wigner slices taken at \textbf{B1} (before SBS) and \textbf{B2} (after SBS).  The green squares at \textbf{B2} for mode~$1$ coincide with the blue crosses representing mode~$2$ at \textbf{B1}, confirming that the cat state originally in mode~$2$ has been transferred to mode~$1$ and rotated by $\pi/2$. The orange circles at \textbf{B2} for mode~$2$ match the orange dashed vacuum curve, demonstrating that mode~$2$ has been reset to vacuum.  These observations quantitatively verify the phonon‑number--dependent phase predicted by Eq.~(\ref{eq:parity-sbs}) and (\ref{eq:motion-phase-rotation}).

\section{Sources of Error \label{sect:error}}

\subsection{Off-Resonant Carrier Excitation \label{sect:error/off-resonant-carrier}}

The bichromatic beams used to implement the spin-dependent beam splitter interaction are symmetrically detuned from the qubit transition by approximately 360 kHz, corresponding to the radial mode splitting. As a result, the carrier transition---with a Rabi frequency on the order of 100 kHz---cannot be neglected during this interaction. The presence of the carrier transition manifests as a rapid oscillation in the spin population during the spin-dependent beam splitter operation.

\begin{figure}
    \centering
    \includegraphics[width=0.4\linewidth]{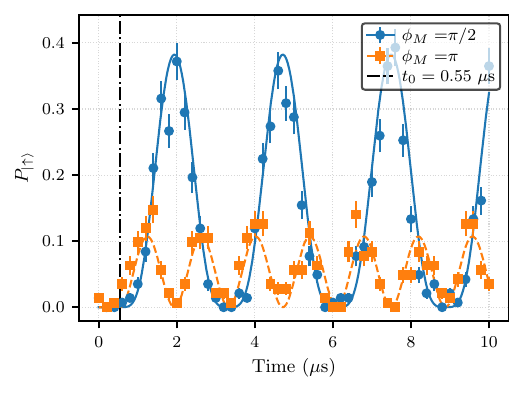}
    \caption{\label{fig:bichromatic-carrier} Time evolution of the spin population under the Hamiltonian in Eq.~(\ref{eq:bichromatic-detuned-carrier}), starting from the initial state $\ket{\downarrow}\ket{0}$. Circle and square markers indicate measured $\ket{\uparrow}$ populations for $\phi_M=\pi/2$ and $\pi$, respectively. The symmetric detuning is $\delta/(2\pi) = 360\,\mathrm{kHz}$, and the Rabi frequency used for the theoretical curves is $\Omega/(2\pi) = 120\,\mathrm{kHz}$. In the experiment, a ramp-up delay of the acousto-optic modulator, $t_0$, is present, indicated by the vertical dash-dot line. For $t < t_0$, the Hamiltonian is set to zero in theoretical curves. The time evolution shows periodicity for both $\phi_M$'s, returning to the initial point $P_{\ket{\uparrow}}=0$ at $t=t_0+n \cdot 2\pi / \delta$, where $n \in \mathbb{N}$. Error bars reflect quantum projection noise, computed from 300 repetitions per data point.}
\end{figure}

A detuned carrier transition can be described by the Hamiltonian
\begin{equation}
    H_\text{carrier} (\delta, \phi) = \frac{\hbar \Omega}{2} (e^{i(\delta t + \phi)} \sigma_+ + e^{-i(\delta t + \phi)} \sigma_-),
\end{equation}
where $\delta$ and $\phi$ are the detuning from the carrier resonance and the phase, respectively. In the case of symmetrically detuned bichromatic beams, the Hamiltonian takes the form
\begin{equation} \label{eq:bichromatic-detuned-carrier}
    H_\text{carrier} (\delta, \phi_b) + H_\text{carrier} (-\delta, \phi_r) = \hbar \Omega \cos (\delta t + \phi_M) \sigma_{\phi_S},
\end{equation}
where $\phi_S$, $\phi_M$, and $\sigma_{\phi_S}$ are defined in the same way as in Eq.~(\ref{eq:ham-sbs}). Despite its time dependence, the Hamiltonian in Eq.~(\ref{eq:bichromatic-detuned-carrier}) commutes with the ideal spin-dependent beam splitter Hamiltonian at all times, as both share the same spin operator $\sigma_{\phi_S}$. As a result, the time-evolution operators corresponding to the beam splitter and the detuned carrier can be separated. Therefore, by choosing the interaction time $t$ as a multiple of $t_\delta\equiv2\pi / \delta$, the effect of the off-resonant carrier transition can be ignored. This is well illustrated in Fig.~\ref{fig:bichromatic-carrier}. This technique is used in all spin-dependent beam splitter (SBS) experiments presented in this manuscript. For example, for each desired SBS interaction time $t$, we round the value to $\tilde t=\lfloor t/t_\delta + 1/2\rfloor t_\delta$ and use it instead of $t$. As $t_\delta\approx2.8\ \mathrm{\mu s}$ for $\delta=360\ \mathrm{kHz}$, it is relatively small adjustment considering the $\pi$-pulse duration of the SBS interaction $t_\pi\approx500\ \mathrm{\mu s}$. 

We note that the spin-dependent beam splitter operation can be realized without the need to factor out the off-resonant carrier dynamics by using a structured beams such as a Laguerre--Gaussian beam and a Hermite--Gaussian beam \cite{Stopp_2022, Peshkov2023Excitation}. Due to the non-Gaussian electric field structure, a Raman transition that does not drive the carrier qubit transition can be implemented. With this, the speed of the beam splitter operation can be increased without being limited by the carrier interaction which is typically two orders of magnitude stronger.

\subsection{Qubit State Detection Fidelity} \label{sect:error/qubit-detection}
We use the threshold discrimination method for qubit state detection. If the number of photon counts registered by the photomultiplier tube (PMT) exceeds a predefined threshold, the state is identified as $\ket{\uparrow}$; otherwise, it is assigned as $\ket{\downarrow}$. The dominant contribution to the detection infidelity arises from erroneously identifying a true $\ket{\uparrow}$ state as $\ket{\downarrow}$, primarily due to off-resonant scattering during the cycling transition between the $S$ and $P$ states. Consequently, a long detection duration can induce leakage of the qubit state, whereas a shorter duration results in insufficient photon counts for reliable state discrimination. We therefore optimize the detection time to balance these effects---ensuring sufficient signal while minimizing qubit state leakage. Nonetheless, the detection process is imperfect. To take this into account, we calibrate the detection error and apply a correction to the measured qubit state populations based on this calibration.

Let $X_\psi$ denote the event that the true qubit state collapses to $\ket{\psi}$ upon detection, and $Y_\psi$ denote the event that the measured qubit state is $\ket{\psi}$, where $\psi \in \left\{\uparrow, \downarrow \right\}$. In our case, the dominant contribution to detection infidelity arises from $P(Y_\downarrow | X_\uparrow)$, which would ideally be zero. Our objective is to estimate the true state probability $P(X_\uparrow)$ from the measured probability $P(Y_\uparrow)$ and the conditional probabilities $P(Y_\uparrow | X_\psi)$. For brevity, let $q_\psi = P(Y_\uparrow | X_\psi)$. Then, by the law of total probability,
\begin{equation}
    P(Y_\uparrow) = q_\uparrow P(X_\uparrow) + q_\downarrow (1 - P(X_\uparrow)).
\end{equation}
Rearranging the terms yields
\begin{equation}
    P(X_\uparrow) = \frac{P(Y_\uparrow) - q_\downarrow}{q_\uparrow - q_\downarrow}.
\end{equation}
Thus, if $q_\uparrow$ and $q_\downarrow$ are known, we can estimate the true qubit population from the measurement data. These conditional probabilities can be calibrated using Rabi oscillations driven by a microwave field resonant with the qubit transition. Given that the qubit coherence time is several hundred milliseconds, we assume that the Rabi oscillation is nearly ideal. By fitting the measured $P(Y_\uparrow)$ to a sinusoidal function, $q_\uparrow$ and $q_\downarrow$ are extracted as the maximum and minimum values of the fitted oscillation, respectively. These values represent the state preparation and measurement (SPAM) fidelity, and in our setup, $q_\uparrow$ and $q_\downarrow$ are typically 0.92--0.96 and 0.00--0.01, respectively.

\subsection{Motional Heating} \label{sect:error/heating-rate}
We identify motional heating as the dominant source of infidelity in the spin-dependent beam splitter interaction in our system. The decay observed in Fig.~\ref{fig:beam-splitter-time-evolution} arises solely from this heating effect. See Appendix~\ref{sect:numerical-simulation} for the numerical simulation method.

\begin{figure}
    \centering
    \includegraphics[width=0.5\linewidth]{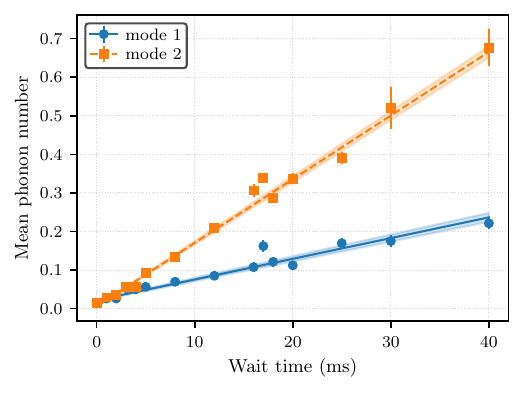}
    \caption{\label{fig:heating-rate} Heating rates of the two radial modes. The fitted heating rates are 5.4(2) and 16.5(3) quanta/s for modes $1$ and $2$, respectively. Each data point is obtained from a simultaneous fit to blue- and red-sideband Rabi oscillations measured over 0--1~ms in 5~$\mu$s step, with 200 repetitions per time point. The error bars denote the fitting uncertainty.}
\end{figure}

Figure~\ref{fig:heating-rate} presents the measured heating rates of our system. These rates are obtained by observing Rabi oscillations on the red and blue sidebands following a variable delay $t_{\text{wait}}$. By scanning $t_\text{wait}$ and assuming that the motional state thermalizes during the delay, we extract the mean phonon number by fitting the oscillation patterns. For each $t_\text{wait}$, we simultaneously fit the red- and blue-sideband oscillations using a single mean phonon number. The exact origin of the heating remains unclear, and the heating rate exhibits day-to-day fluctuations. The heating rates shown in Fig.~\ref{fig:heating-rate} correspond to the beam splitter time evolution data in Fig.~\ref{fig:beam-splitter-time-evolution}. For the two-mode Wigner function measurements in Figs.~\ref{fig:wigner-fock}, \ref{fig:wigner-ecs-even}, and \ref{fig:wigner-ecs-odd}, the heating rates for modes~$1$ and $2$ measured on the day of the experiment are 6.1 and 39.0 quanta/s, respectively.

\subsection{Finite Motional Coherence Time} \label{sect:error/finite-motional-coherence-time}
The stability of the motional mode frequencies plays a critical role in determining the fidelity of spin-dependent forces. Frequency noise in the motional modes detunes the spin-dependent force from resonance and causes fluctuations in its direction. For example, in the Wigner function measurement experiment, we apply spin-dependent forces sequentially to each mode and perform spin-state measurement to generate an entangled coherent state. This process includes a mid-circuit detection duration of 250 $\mathrm{\mu s}$. Following this, additional spin-dependent forces are applied, along with carrier $\pi/2$-pulses, to displace the generated ECS. Any variation in the mode frequencies over this timescale introduces an error in the final motional state.

Moreover, if the mode frequency varies slowly compared to the timescale of a single experimental shot, the error accumulates coherently during the mid-circuit detection process. As a result, the direction of subsequent spin-dependent forces deviates significantly from the intended trajectory. Such shot-to-shot variations cannot be accurately captured by standard Lindblad master equations, as the noise is inhomogeneous across experimental repetitions.

\begin{figure}
    \centering
    \subfloat[]{
        \includegraphics[width=0.4\linewidth]{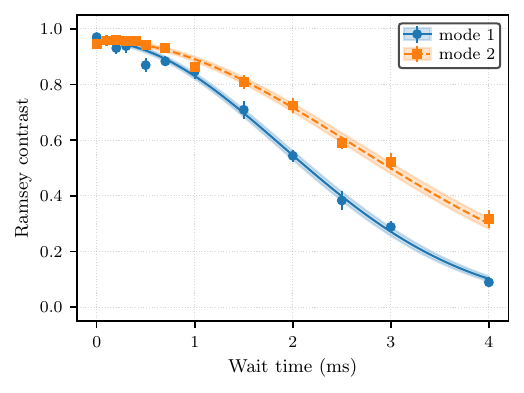}
    }
    \subfloat[]{
        \includegraphics[width=0.4\linewidth]{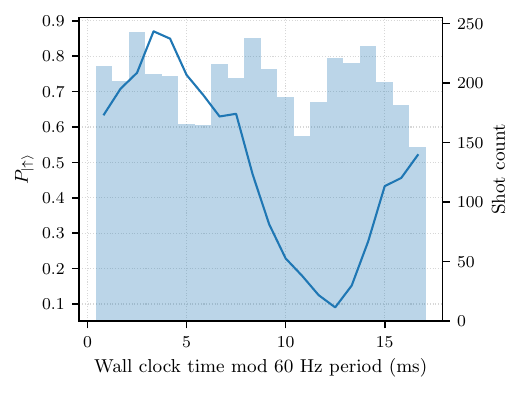}
    }
    \caption{\label{fig:single-mode-coherence} (a) Decay of single-mode blue-sideband Ramsey contrast. Scatter markers with error bars represent the fitted Ramsey contrast from experimental data, measured over phases 0--$4\pi$ in steps of $\pi/9$, with 150 repetitions per phase point. Solid and dashed curves show the Gaussian decay fits. See main text for details. (b) 60-Hz modulation of the motional frequency of mode~$1$. A blue sideband Ramsey experiment with a waiting time $t_\text{wait}$ of 2 ms is repeated, and the result is reorganized based on the absolute timestamp recorded for each run. The horizontal axis shows the timestamp modulo $1/60$ seconds. The solid curve represents the measured spin population and semi-transparent bars indicate the number of experimental shots in each modulo-time bin.}
\end{figure}

To characterize the motional coherence time, we measured the Ramsey contrast of the blue-sideband transition between $\ket{\downarrow}\ket{0}$ and $\ket{\uparrow}\ket{1}$ as shown in Fig.~\ref{fig:single-mode-coherence}(a). For each wait time $t_\text{wait}$, we scan the phase of the second $\pi/2$-pulse, from $0$ to $2\pi$, resulting in a single period of sinusoidal oscillation. We fit the data to extract the oscillation amplitude (the Ramsey contrast) and its uncertainty. Since the spin coherence time exceeds several hundred milliseconds, the observed decay in Ramsey contrast is attributed primarily to motional decoherence. Although heating can also contribute to dephasing, the effect is negligible given the measured heating rate.

If the motional decoherence can be modeled using Lindblad collapse operators $a_j^\dagger a_j$, the Ramsey contrast is expected to decay exponentially as $e^{-\gamma t}$. However, the experimental data in Fig.~\ref{fig:single-mode-coherence}(a) more closely follow a Gaussian decay of the form $e^{-\gamma t^2}$ as shown by the fitted curves. Several models predict this behavior, including decoherence from shot-by-shot Gaussian-distributed noise. These observations suggest that the dominant source of motional decoherence arises from inhomogeneous mode-frequency fluctuations across experimental shots. Indeed, Fig.~\ref{fig:single-mode-coherence}(b) clearly shows that the mode frequency is modulated at 60 Hz. From the oscillation amplitude of the spin population, we estimate the peak-to-peak amplitude of the mode-frequency noise to be approximately 140 Hz. We suspect that the trap electronics pick up the mains power, which is 60-Hz AC. The power of the trap RF is sampled via a capacitive voltage divider, and is stabilized by a servo controller, whose control output is fed into an RF mixer.

The spin-dependent beam splitter (SBS) interaction is affected by motional decoherence. In particular, the stability of the difference between the two motional modes is critical for maintaining the coherence of the SBS operation. A separate experiment found that the fluctuations of the individual mode frequencies exhibit a positive correlation, indicating that the differential mode frequency remains more stable than each frequency independently. As a result, the effective coherence time for the beam splitter interaction exceeds that of the individual modes.

\begin{figure}
    \centering
    \subfloat[]{
        \includegraphics[width=0.4\linewidth]{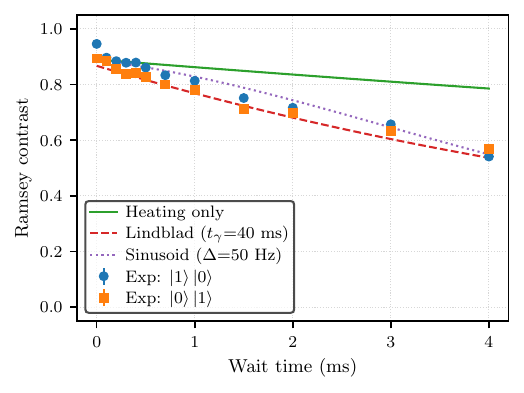}
    }
    \subfloat[]{
        \includegraphics[width=0.4\linewidth]{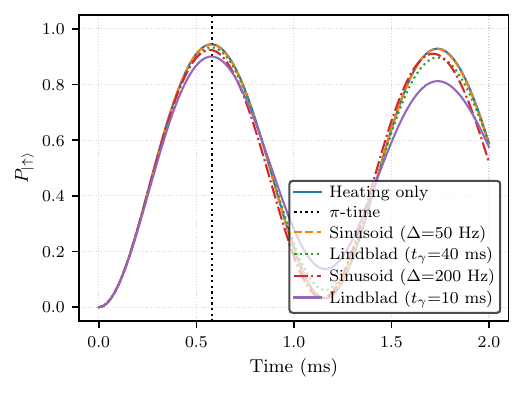}
    }
    \caption{\label{fig:beam-splitter-coherence} (a) Decay of beam splitter Ramsey contrast. Scatter markers with error bars represent the fitted Ramsey contrast from experimental data, measured over phases 0--$4\pi$ in steps of $\pi/18$, with 150 repetitions per phase point. The three curves show numerical simulations under different noise conditions. See main text for details. (b) Numerical simulations of the time evolution of spin population under the spin-dependent beam splitter interaction, starting from the initial state $\ket{1}\ket{0}$, with various noise conditions.}
\end{figure}

To demonstrate the motional coherence of the SBS interaction, we perform another Ramsey-type experiment analogous to the single-mode case, but using SBS $\pi/2$-pulses instead of blue-sideband pulses. The result shown in Fig.~\ref{fig:beam-splitter-coherence}(a) suggests the presence of motional decoherence beyond heating, with a trend resembling Lindbladian dephasing. Notably, the observed decay is significantly slower than that of the individual single-mode experiments. Figure~\ref{fig:beam-splitter-coherence}(b) shows that the decoherence rate is comparatively slow, and that the contrast of the spin population under the SBS interaction is largely preserved. The noise models ``Lindblad($t_\gamma$)'' refers to a Lindblad master equation with collapse operator $(a_1^\dagger a_1 - a_2^\dagger a_2)$ and coherence time $t_\gamma$, while ``Sinusoid($\Delta$)'' refers to a 60-Hz sinusoidal fluctuation in the differential mode frequency $(\omega_2 - \omega_1)$ with amplitude $\Delta$, respectively. Motional heating is included in both models.

\subsection{Lamb--Dicke Approximation \label{sect:error/lamb-dicke-approximation}}
In implementing the spin-dependent force and the spin-dependent beam splitter Hamiltonian, we consider only the dominant lowest-order terms. This approximation is justified within the Lamb--Dicke regime, where $\eta^2n \ll 1$, with $\eta$ denoting the Lamb--Dicke parameter and $n$ the phonon number. However, higher-order terms introduce small corrections to the interactions. This effect is prominent in most of the interactions between the ion and the laser. For instance, a decay in carrier Rabi oscillations is observed when the system is prepared in a superposition or mixture of multiple Fock states, indicating that residual higher-order terms are not entirely negligible.

To simulate the effect of higher-order terms on the spin-dependent beam splitter interaction, we compute the matrix exponential $\exp \left( i \eta (a_j^\dagger + a_j) \right)$ to model the dynamics exactly, without invoking the Lamb--Dicke approximation (see Appendix~\ref{sect:numerical-simulation}). Figure~\ref{fig:lda}(a) illustrates the breakdown of the Lamb--Dicke approximation for the spin-dependent force when the displacement becomes large. In this simulation, the Hilbert space dimension of the motional mode is set to 200, and the Lamb--Dicke parameter is 0.1. The slower increase in mean phonon number clearly indicates that the spin-dependent force deviates from the ideal displacement operator for large $\alpha$. This deviation not only reduces the displacement amplitude but also distorts the quantum state, as evidenced by the fidelity between the generated state and the ideal coherent state $\ket{\sqrt{\bar n}}$ being less than unity. In our experiment, however, this effect remains negligible, as the effective displacement amplitudes reach at most approximately 4---in case where an entangled coherent state with $\alpha_1=\alpha_2=1.2$ is displaced by $\beta_1=\beta_2=\pm2.5$.

The Lamb--Dicke approximation can play a critical role in the spin-dependent beam splitter interaction when applied to largely displaced states, as illustrated in Fig.~\ref{fig:lda}(b). As the phonon number increases, the $\pi$-time, where the joint-parity is mapped to the spin state, is pushed back. For our maximum displacement amplitude---approximately 4---the corresponding mean phonon number reaches 16, and this breakdown of approximation can introduce significant error, since the $\pi$-pulse duration is always calibrated using the $\ket{1}\ket{0}$ state. However, in most cases, the key features of interest---such as the interference patterns in the Wigner function of entangled coherent states---are determined primarily by low-lying Fock states. In regions far from the origin of phase space, the expected Wigner function approaches zero, implying that even and odd Fock numbers are equally populated. In such cases, neighboring Fock states contribute with opposite sign to the spin population and effectively cancel each other, making the deviation from ideal dynamics negligible.

\begin{figure}
    \centering
    \subfloat[]{
        \includegraphics[width=0.4\textwidth]{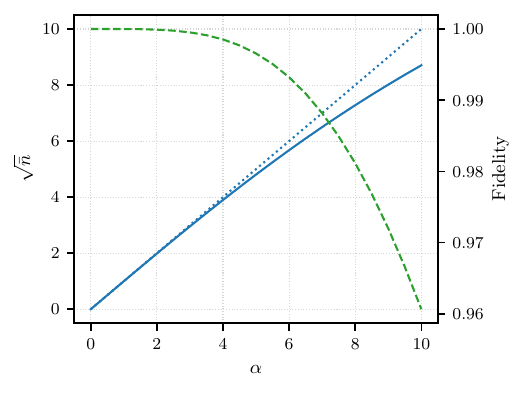}
    }
    \subfloat[]{
        \includegraphics[width=0.4\textwidth]{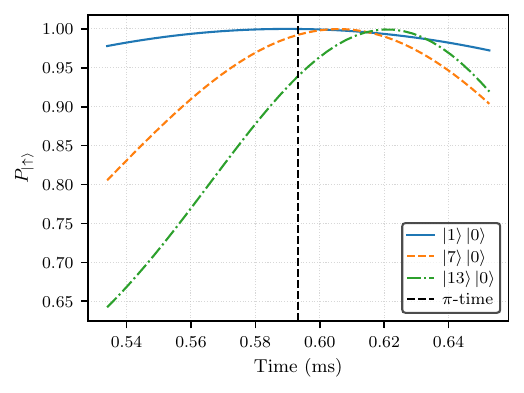}
    }
    \caption{\label{fig:lda} Breakdown of Lamb--Dicke approximation (LDA). (a) Fidelity degradation of coherent states generated by a spin-dependent force, without applying the LDA, obtained via numerical simulation. The dotted diagonal line indicates the relation $\bar n = | \alpha |^2$ for ideal coherent states $\ket\alpha$, where $\bar n$ is the mean phonon number. The solid curve shows that the increment in mean phonon number slows as $\alpha$ increases. The dashed curve illustrates the decrease in fidelity between the generated state and an ideal coherent state $\ket{\sqrt{\bar n}}$. (b) Numerically simulated time evolution of the spin population under the spin-dependent beam splitter interaction, wihtout applying the LDA, for various Fock states. The vertical dashed line indicates the ``standard'' $\pi$-time, calibrated using the $\ket{1}\ket{0}$ state.}
\end{figure}

\section{Calibration Routine} \label{sect:calibration-routine}
To achieve high fidelity operations, the experimental parameters must be precisely calibrated. In the experiment, we employ four types of interactions: the carrier transition, the sideband transitions, the spin-dependent forces (SDFs), and the spin-dependent beam splitter (SBS). Each of these interactions requires careful calibration to ensure consistent and accurate quantum control.

For state detection, we observe Rabi oscillations of the spin state under a resonant microwave drive. As this operation is highly reliable, we fit this oscillation to extract $q_\uparrow$ and $q_\downarrow$, as defined in Appendix~\ref{sect:error/qubit-detection}. The primary sources of fluctuation in these values are drifts in the optical frequency and power of the detection beam, which results from drift in the wavelength meter used for frequency stabilization and from beam alignment drift, respectively.

For the carrier transition, the Rabi frequency is calibrated to 100 kHz by adjusting the amplitude of the RF signal driving the acousto-optic modulator (AOM). The amplitude is tuned to minimize the spin population at a pulse duration of 50 $\mu$s, corresponding to the fifth period of a 100 kHz Rabi oscillation. This calibrated amplitude is then used as a reference to determine the amplitudes for other interactions.

For the interactions involving bichromatic laser beams, such as SDF and SBS, intensity balance is critical. In the current setup, the bichromatic beams become spatially separated at the ion location due to the use of an AOM without relay lenses. As a result, the intensity balance is highly sensitive to beam alignment. To mitigate this, we employ motorized stages to automatically align the beams by adjusting the positions of the focusing lenses. This alignment is performed by scanning the stage positions, measuring the beam profiles, and determining the center via Gaussian fitting. Since the spatial separation scales with frequency detuning, the SDF interaction is more susceptible to this effect than the SBS. To compensate for residual imbalance, we calibrate the amplitude ratio between the red- and blue-detuned RF tones that drive the AOM for the SDF.

To obtain the radial mode splitting for resonant SDF, we measure the individual mode frequencies via a Ramsey-type experiment using the SDF. Starting from the initial state $\ket{\downarrow}\ket{0}$, we apply the SDF to generate the entangled state $\ket{+}\ket{\alpha} + \ket{-}\ket{-\alpha}$ (normalization omitted). After a waiting period, a second SDF pulse of equal duration and opposite motional phase is applied. If the interaction is on resonance, the system returns to the initial state, leaving the spin in $\ket{\downarrow}$. Otherwise, the force direction rotates during the wait time, preventing full disentanglement between spin and motion, and resulting in a spin population near 0.5. Therefore, we scan the symmetric detuning of the SDF and fit the resulting spin populations to the theoretical curve to extract the center of the resonance dip.

Since the SBS interaction is slow and has a narrow transition line---on the order of 1 kHz---, it is essential to precisely determine the laser detunings required to drive the interaction. The symmetric detuning can be readily determined from the measured mode frequency difference. However, we observe a fluctuating $\sigma_z$ term in the actual Hamiltonian, likely arising from residual AC Stark shifts. To mitigate this, we introduce an offset detuning---applied equally to both bichromatic tones---which cancels the $\sigma_z$ term in the interaction frame. We extract this offset by measuring the resonance conditions for the initial states $\ket{1}\ket{0}$ and $\ket{0}\ket{1}$ and determining the detuning that symmetrizes the response.

The calibration routines described above are fully automated and performed approximately every 20 minutes, each taking 6--8 minutes, resulting in an effective experimental duty cycle of about 60--70\%.

\section{Numerical Simulation \label{sect:numerical-simulation}}

For numerical simulations, we use version 4.7.6 of the Quantum Toolbox in Python (QuTiP)~\cite{johansson_qutip_2012}. Since all Hamiltonians in this manuscript are implemented using Raman beams, the simulations are also constructed from detuned or resonant Raman transition Hamiltonians. For instance, we define such a Hamiltonian as ``single-beam'' as follows:
\begin{equation} \label{eq:single-beam-hamiltonian}
    H_\text{single}^{(n_1, n_2)}(t ; \Omega, \phi) = \frac{\Omega}{2} e^{i\phi} \sigma_+ \, \operatorname{diag}_{-n_1} \left( e ^ {i \eta_1 \left( a_1^\dagger + a_1 \right)} \right) \, \operatorname{diag}_{-n_2} \left( e ^ {i \eta_2 \left( a_2^\dagger + a_2 \right)} \right) + \text{H.c.},
\end{equation}
where $\operatorname{diag}_k (A)$ refers to the $k$-th diagonal operator of a linear operator $A$, whose matrix elements are defined by
\begin{equation}
    \left[ \operatorname{diag}_k (A) \right]_{mn} = 
    \begin{cases}
        A_{mn}, & \text{if } m - n = k, \\
        0, & \text{otherwise.}
    \end{cases}
\end{equation}
In Eq.~(\ref{eq:single-beam-hamiltonian}), $n_1$ and $n_2$ denote the sideband orders of the transition. For example, the superscript $(1, -1)$ indicates a red-detuned beam-splitting sideband (recall that $\omega_1 < \omega_2$) that creates a phonon in mode~$1$, removes a phonon from mode~$2$, and flips the spin from $\ket{\downarrow}$ to $\ket{\uparrow}$. $\Omega$ and $\phi$ represent Rabi frequency and phase, respectively. In the presence of a detuning $\delta$, its effect can be incorporated by allowing the phase $\phi$ to become time-dependent, $\phi(t)=\delta t + \phi_0$. The carrier transition Hamiltonian corresponds to $H_\text{single}^{(0, 0)}$, while the blue and red sidebands of mode $a$ are given by $H_\text{single}^{(\pm1, 0)}$. The spin-dependent force in mode $a$ and the spin-dependent beam splitter interaction can be written as $H_\text{single}^{(-1, 0)} + H_\text{single}^{(1, 0)}$ and $H_\text{single}^{(1, -1)} + H_\text{single}^{(-1, 1)}$, respectively, with appropriate choices of the phase parameter $\phi$.

By construction, the Hamiltonian $H_\text{single}$ does not invoke the Lamb--Dicke approximation. Although the breakdown of the approximation is largely negligible in our experiments (see Sec.~\ref{sect:error/lamb-dicke-approximation}), all simulations in this manuscript use $H_\text{single}$ without assuming the approximation. The Lamb--Dicke parameters used in the simulations are 0.1 and 0.087 for modes~$1$ and $2$, respectively. The Hilbert space dimensions of the motional modes are set to 20 unless otherwise specified.

To simulate motional heating, we numerically solve the Lindblad master equation using the QuTiP package, incorporating collapse operators $a_j$ and $a_j^\dagger$~\cite{Intravaia2003Quantum}. In our case, as the temperature of the reservoir is the room temperature, we use equal rates for $a_j$ and $a_j^\dagger$, corresponding to $\Gamma (N+1)$ and $\Gamma N$ in Ref.~\cite{Intravaia2003Quantum}, respectively.

The 60-Hz fluctuation of the mode frequencies is simulated by modulating the motional phase $\phi_M$ in the spin-dependent force Hamiltonian, Eq.~(\ref{eq:hamiltonian-sdf}). To implement this spin-dependent force for mode~$1$, for example, we use bichromatic beams that are symmetrically detuned from the carrier transition by the mode frequency $\omega_1$. In this case, each beam becomes resonant, yielding the Hamiltonian $H_\text{single}^{(1, 0)}(t ; \Omega, \phi_b) + H_\text{single}^{(-1, 0)}(t ; \Omega, \phi_r)$. However, if the mode frequency deviates from the symmetric detuning by a constant $\delta$, residual detunings are introduced into the Hamiltonian:
\begin{equation} \label{eq:detuned-sdf}
    H_\text{single}^{(1, 0)}(t ; \Omega, -\delta t + \phi_b) + H_\text{single}^{(-1, 0)}(t ; \Omega, \delta t + \phi_r).
\end{equation}
As a result, the motional phase of the spin-dependent force becomes time-dependent:
\begin{equation} \label{eq:motional-phase-detuned-sdf}
    \phi_M(t)=(-2\delta t + \phi_b - \phi_r) / 2=-\delta t + \phi_{M, 0}.
\end{equation}
In the presence of 60-Hz noise, the $\delta$ becomes time-dependent and oscillates at 60 Hz:
\begin{equation}
    \delta(t) = \Delta \sin (\omega_\text{noise}t + \phi_\text{noise}),
\end{equation}
where $\omega_\text{noise} / (2\pi)$ is 60 Hz, and $\Delta$ and $\phi_\text{noise}$ denote the amplitude and phase of the detuning oscillation. Since $\delta(t)$ is time-dependent, the linear $\delta t$ terms in Eqs.~(\ref{eq:detuned-sdf}) and (\ref{eq:motional-phase-detuned-sdf}) must be replaced by the time integral $\int_0^t {\delta (\tau) d \tau}$, yielding:
\begin{equation}
    \phi_M(t) = \frac{\Delta}{\omega_\text{noise}} \left( \cos (\omega_\text{noise}t + \phi_\text{noise}) - \cos \phi_\text{noise} \right) + \phi_{M,0}.
\end{equation}

This phase modulation is applied coherently within each experimental run. For example, in the Wigner function measurement of an entangled coherent state, we first displace the vacuum state to generate the ECS, and then apply a second displacement to perform the measurement. Although the spin-dependent force interaction is turned off between the two displacement operations, the motional phase continues to accumulate during this interval, ultimately determining the direction of the second force. However, the phase of the noise, $\phi_\text{noise}$, is randomly set for each experimental shot, resulting in a decoherence effect. To simulate this behavior, we repeat the simulation for eight evenly spaced values of $\phi_\text{noise}$ (i.e., $0$, $\pi/4$, $\pi/2$, \ldots, $7\pi/4$) and take the average. The simulations shown in Fig.~\ref{fig:wigner-ecs-even} use a noise amplitude of $\Delta / (2\pi) = 150$ Hz.

\bibliography{ref3}

\end{document}